\documentstyle[aps,prb]{revtex}
\begin{document}
\thispagestyle{empty}
\pagestyle{plain}
\draft
\title{Analysis of the valence band photoemission spectrum of 
Sr$_2$CuO$_2$Cl$_2$ along
the high-symmetry directions}
\author{R. Hayn, and H. Rosner}
\address{Institute for Theoretical Physics, TU Dresden, 01062
Dresden, Germany}
\author{V. Yu. Yushankhai}
\address{Joint Institute for Nuclear Research, 141980 Dubna, Moscow 
Region,
Russia}
\author{S. Haffner, C. D\"urr, M. Knupfer, G. Krabbes, M. S. Golden, 
J. Fink, and H. Eschrig}
\address{Institute for Solid State and Materials Research (IFW) 
Dresden,
\\P. O. Box 270016, D-01171 Dresden, Germany}
\author{D. J. Singh}
\address{Complex Systems Theory Branch, Naval Research Laboratory, 
Washington,
DC 20375-8795, USA}
\author{N.T. Hien, and A.A. Menovsky}
\address{Van der Waals-Zeeman Laboratory, University of Amsterdam,
Valckenierstraat 65, 1018 XE Amsterdam}
\author{Ch. Jung, G. Reichardt}
\address{BESSY GmbH, Lentzeallee 100, D-14195 Berlin, Germany}

\date{\today}

\maketitle

\begin{abstract}

Band structure calculations have been used to identify the different bands
contributing to the polarisation-dependent photoemission spectra
of the undoped model cuprate Sr$_2$CuO$_2$Cl$_2$ at the
high-symmetry points of the CuO$_2$ plane $\Gamma$, $(\pi/a,0)$ and 
$(\pi/a,\pi/a)$ and 
along the high-symmetry directions $\Gamma - (\pi/a,\pi/a)$ and $\Gamma 
- (\pi/a,0)$.
Results from calculations within the local density approximation (LDA) have
been  
compared with 
calculations taking into account the strong electron correlations by 
LDA+U, with the result that 
the experimental order of energy levels at the high-symmetry points is 
better described by the LDA+U
calculation than by the simple LDA.
All the main peaks in the photoemission spectra at the high symmetry 
points could be assigned to different
Cu 3$d$ and O 2$p$ orbitals which we have classified according to their 
point symmetries. 
The dispersions along the high-symmetry directions were compared with 
an 11-band 
tight-binding model which was fitted both to the LDA+U band structure
calculation and the angle-resolved photoemission data. The mean field 
treatment
successfully describes the oxygen derived bands but shows discrepancies 
for the 
copper ones. \\[0.5cm]

PACS-numbers: 74.25 Jb, 74.72 Jt, 79.60 Bm 

\end{abstract}

\section{Introduction}
One strategy to answer the many questions concerning the electronic  
structure
of cuprate superconductors is the study of model substances. One of  
these
compounds is Sr$_2$CuO$_2$Cl$_2$. It
is a two dimensional (2D) antiferromagnetic (AFM) insulator with a N\'eel
temperature \cite{geven} of 256 K  whose magnetic structure is well  
described
by the 2D spin
1/2 Heisenberg-model. It was the first undoped cuprate which allowed  
the 
angle
resolved photoemission (ARPES) measurement of its lowest excitations.
\cite{wells,larosa,kim} These excitations are well described by one  
hole in a
2D quantum
antiferromagnet. \cite{horsch,dagotto,becker} Deviations from the  
one-hole
dispersion of the pure $t$-$J$ model can be reduced by taking into  
account
hopping terms to second and third neighbors.  
\cite{nazarenko,yushankhai} In
the same substance the low binding energy edge of the main valence
band has been interpreted in terms of 
non-bonding oxygen orbitals which are completely decoupled from the  
copper
system. \cite{pothuizen} (These features were known before
as ``1 eV-peak''.) \cite{tobin} These non-bonding states are especially
pronounced at $(\pi,\pi)$ where they have minimal binding energy. But  
the
detailed structure of the complete valence band has never been analyzed  
up to
now and that is the aim of the present work. Furthermore, we will show  
that
one can obtain additional information on the low binding-energy  
features by
analyzing their dependence on the polarization of the photon.

Polarization dependent photoemission measurements are an effective tool  
to
analyze the electronic structure of the valence band in detail. By  
measuring
along high-symmetry directions all bands can be classified according to  
their
symmetry properties. This allows a very precise comparison between  
experiment
and different theoretical predictions. It is well established that in  
all the
cuprates electron correlations have a strong  influence on the  
electronic bands near
the Fermi level which is especially pronounced in undoped substances.  
\cite{dagotto}
But the influence of correlations on those parts of the valence band  
with larger
binding energies is less clear. We will show that the combination of
polarization dependent ARPES measurements with theoretical  
investigations
taking into account the electron correlation to a differing extent  
(LDA, LDA+U)
provides a unique possibility to answer this question.

The model cuprate Sr$_2$CuO$_2$Cl$_2$ is very well suited for such an
investigation. It has a tetragonal structure with ideal planar
CuO$_2$ layers \cite{miller} and cleaves readily parallel to the CuO$_2$  
planes.
Furthermore, the presence of Cl instead of apex
oxygen allows a restriction of the states which contribute to the ARPES  
spectra to those of the pure
CuO$_2$ plane alone.
This can be achieved by choosing a photon energy close to the Cooper  
minimum
for Cl 3$p$ photoemission, i.e. Cl 3$p$ states will then have a small  
photon cross section. In this manner
we intend to study an ideal situation whose main characteristics should  
be generic to all the cuprates.

Recently, a similar study was presented for the Cu$_3$O$_4$ plane of
Ba$_2$Cu$_3$O$_4$Cl$_2$. \cite{schmelz} It turns out that  
Sr$_2$CuO$_2$Cl$_2$
is considerably less complex than Ba$_2$Cu$_3$O$_4$Cl$_2$. Therefore,  
we are now able to
identify {\em all} the peaks at the high symmetry points in contrast to
Ba$_2$Cu$_3$O$_4$Cl$_2$ where only the upper parts of the valence band  
were
analyzed.
This allows implications about the influence of electron correlations  
on the
valence band (VB) structure of Sr$_2$CuO$_2$Cl$_2$ to be made. It is  
known that simple LDA fails to predict the insulating ground state of undoped
cuprates. \cite{pickett} There are several improvements of LDA such as  
the
self-interaction correction (SIC) method \cite{perdew} or LDA+U
\cite{anisimov} which has already been applied to the case of  
lanthanum cuprate \cite{czyzyk} (La$_2$CuO$_4$). 
Here, we apply LDA+U plus a symmetry analysis at special $k$-points to  
interprete
the polarization dependent photoemission data for Sr$_2$CuO$_2$Cl$_2$,  
where
the actual value of U is chosen to describe the experimental situation.

The paper is organized as follows. After describing the experimental  
method 
and
the details of the LDA band structure calculation we analyze the  
symmetry
properties of the wave function along high-symmetry directions. The
symmetry properties of the relevant bands are most clearly seen in a
tight-binding model presented in Sec.\ IV. Discussing the correlation  
effects
in a mean-field manner leads us to an LDA+U calculation whose results  
are
presented in Sec.\ V. In Sec.\ VI we compare the experimental spectra  
with 
the
theoretical predictions. The spectra at $\Gamma$, $(\pi,0)$ and  
$(\pi,\pi)$
(the lattice constant has been set to unity in all the notations) can  
be
understood from the LDA+U but not from the LDA calculation. The
experimental dispersion relations are discussed in terms of a  
tight-binding (TB)-model.

\section{Experimental}

The Sr$_2$CuO$_2$Cl$_2$ single crystals were grown from the melt, their
typical dimensions being 3 $\times$ 3 $\times$ 0.5 mm. The crystals  
were mounted on
the sample holders using conducting, or in some cases, insulating  
epoxy. If
insulating epoxy was used, electrical contact between sample holder and
sample was achieved by means of a graphite layer at the sides of the  
crystal.
The orientation of the single crystals was determined ex-situ by x-ray  
diffraction. The surface
normal of the crystals is perpendicular to the CuO$_2$ planes. Prior to  
the ARPES
measurements, a clean crystal surface was prepared in ultra high vacuum
(UHV)
by stripping off an adhesive tape which was attached on the sample
surface.

The ARPES measurements were performed using linearly polarized 35 eV
photons
from the crossed undulator beamline U2 of the BESSY I facility and 
BESSY's HIRES photoelectron spectrometer. \cite{Peatman} The angular
resolution was set to $\pm 1^{\circ}$ which gives a momentum resolution
of $\pm$ 0.05 \AA$^{-1}$ for states of 1 eV binding energy, this  
corresponds
to 12 \% of the distance between $\Gamma$ and ($\pi$,0). A total energy
resolution (resulting from both the monochromator and electron analyzer
resolutions) of 150 meV was applied. The electron analyzer is fixed in  
the horizontal plane at an angle of
60$^{\circ}$ with respect to the incoming photon beam, i.e. the  
emission plane which
is defined to be the plane spanned by the sample surface normal and the
$k$ vectors of the measured photoelectrons, is always a horizontal
plane. The photoelectron momentum vector could then be changed by  
variation
of the angle between the axis of the input lens of the electron  
analyzer
and the sample surface normal by rotation of the manipulator on which
the
sample holder was mounted. The ARPES spectra have been recorded in the
electron distribution curve (EDC) mode from both $\Gamma$ to
($\pi$,$\pi$)
and $\Gamma$ to ($\pi$,0). In the latter case, the sample was oriented
in such a way that the Cu - O
bonds of the CuO$_2$ plane were aligned parallel to the fixed,
horizontal
emission plane, while in the former case, the Cu - O bonds were aligned
at an angle of 45$^{\circ}$ with respect to the emission plane. The
spectra are normalized with respect to the incoming photon flux
which was simultaneously measured using the drain current of
a gold mesh. The absolute binding energy scale was determined according
to
$E_B = h \nu	- E_{kin} - \phi_{analyzer}$ using the photon energy,  
the
measured
kinetic energy of the photoelectrons and the known analyzer work
function. There were no indications of charging-induced energy shifts.
All spectra were recorded at 300 K within 8 hours of
a cleave, during which time the samples showed no indications of
surface degradation.

The electronic structure of the VB is derived from O 2$p$,
Cu 3$d$ and Cl 3$p$ orbitals, but for 35 eV photon energy,
the photoionization cross section of the Cl 3$p$ orbitals is much
smaller than that of the O 2$p$ and Cu 3$d$ orbitals \cite{Yeh},
which therefore dominate the ARPES VB spectra.
The measurements had been performed at room temperature which is slightly
above the  
Ne\'el
temperature of Sr$_2$CuO$_2$Cl$_2$ (256 K). Although we are aware that AFM
fluctuations are important, we nevertheless analyzed the spectra 
in terms of the first Brillouin zone (BZ)
of the paramagnetic CuO$_2$
plane of Sr$_2$CuO$_2$Cl$_2$. We will see that this is especially  
justified
for the bands with dominant oxygen character, whereas one observes some
deviations for those bands which couple strongly with the copper spins.
The geometrical structure of a CuO$_2$ plane has two mirror planes
(denoted M$_1$ and M$_2$ in Fig.\ 1). \cite{remarkmp}
All the bands with a wave vector between $\Gamma$
and $(\pi,\pi)$, parallel to the mirror plane M$_1$, can be classified  
to be
either symmetric or antisymmetric with respect to M$_1$, and  
analogously also
for wave vectors along $\Gamma-(\pi,0)$ with respect to reflections
at M$_2$. Experimental information
about the parity of the valence band states with respect
to a mirror plane can be obtained by recording the ARPES spectra with
either perpendicular or parallel polarization of the
electric field vector of the incoming radiation with respect to an
emission plane which is parallel to a mirror plane of the
system. It can then be shown \cite{Zangwill} that for parallel
polarization only initial valence band states which are even with  
respect to 
the
emission/mirror plane contribute to an ARPES spectrum while for
perpendicular polarization, only states
which are odd with respect to the emission/mirror plane are seen in
a spectrum. In this work, the polarization of the electric field vector
was chosen by using either the vertical or horizontal undulator, which
corresponds to perpendicular and parallel polarization with respect to
the emission plane. The emission plane is parallel to mirror plane
M$_1$,
if the ARPES spectra are recorded along the ($\pi$,$\pi$) direction,
while it is parallel to the mirror plane M$_2$ for spectra along the
($\pi$,0) direction. For perpendicular polarization, the electric field
vector is always parallel to the CuO$_2$ planes, i.e. only in-plane orbitals
as O 2$p_{x,y}$ or Cu 3$d_{x^2-y^2}$ contribute to the spectra. For
parallel polarization, the electric field vector is completely in-plane
only at normal-incidence, at any other incidence angle of the photon
beam,
the electric field vector has an out-of-plane component and there are
also
contributions from out-of-plane orbitals such as O 2$p_z$ to the ARPES
spectra.

\section{Band structure calculations}
Sr$_2$CuO$_2$Cl$_2$ has a body centered
tetragonal crystal structure with the lattice
constants $a=3.973$ \AA \
and $c=15.618$ \AA. \cite{miller} Band structure calculations have  
been
performed
treating the exchange and correlation potential within the local  
density
approximation (LDA). The Bloch wave functions were constructed from
atomic-like wavefunctions according to the linear combination of atomic
orbitals (LCAO) method. The calculation was scalar relativistic and due  
to 
the open crystal structure two empty spheres per elementary cell were
introduced in between two oxygen atoms of neighboring CuO$_2$ planes. A
minimal basis was chosen consisting of   
Sr(5$s$,5$p$,4$d$),
Cu(4$s$,4$p$,3$d$), O(2$s$,2$p$), Cl(3$s$,3$p$) orbitals and the 1$s$ and 2$p$
orbitals for the empty spheres. To  
optimize the
local basis a contraction potential $(r/r_0)^4$ was introduced.
\cite{eschriglcao} The Coulomb potential is constructed as a sum of
overlapping  contributions of spherical symmetry and for the exchange  
and
correlation potential the atomic sphere approximation (ASA) is used.

In the resulting band structure (Fig.\ 2) one observes an antibonding  
band
built up of Cu 3$d_{x^2-y^2}$ and O 2$p_{x,y}$ orbitals crossing the  
Fermi
level. This contradicts the experimentally observed non-metallic  
behavior
which already indicates that one has to treat the electron
correlations in a more explicit way. One could conjecture that the only  
effect
of correlations is to split the half-filled antibonding band leaving  
the
structure of the other valence bands roughly unchanged. That is not the  
case,
however, as will become clear from our following analysis. One can also  
observe
in Fig.\ 2 that there is nearly no dispersion of the relevant band in the $z$
direction and 
all discussions in the present paper will be restricted to the
CuO$_2$ plane only.

To check the minimal basis LCAO method,
a full potential linearized augmented plane wave (FLAPW) calculation  
has been
performed for Sr$_2$CuO$_2$Cl$_2$ (Fig.\ 3, compare also
Ref. \onlinecite{freeman}). This method involves no shape approximations and
uses a flexible basis in all regions of space. \cite{singh2} As such it is
well suited to 
open structures with low site symmetries as in the present cuprate. 
We note a sufficiently good agreement  
between both band structures,
although the LCAO-bandwidth of the valence band is found to be somewhat  
larger.
However, there are no significant differences in the order of energy  
levels
between LCAO-LDA and FLAPW-LDA. We stick to the LCAO because we want to
exploit the minimum basis orbital analysis. 

To obtain more information about the structure of the valence band in
our LCAO-LDA we have calculated the orbital weight (defined in Ref.  
\onlinecite{rosnertb})
of each band at the high symmetry points. Due to the low cross section  
of the Cl 3$p$ orbitals for 35 eV
photon energy we concentrate on the Cu 3$d$ and O 2$p$ orbitals (i.e.\  
on
11 bands). The eigenfunctions with a dominant contribution from Cu 3$d$  
and O 2$p$ orbitals
are collected in Table 1. The in-plane oxygen orbitals are divided into  
$p_{\sigma}$ orbitals
which are directed to the Cu site and $p_{\pi}$ orbitals perpendicular
to them.  \cite{mattheiss}
There are two combinations for each: $p_{\sigma}$ and $\tilde  
p_{\sigma}$,
($p_{\pi}$ and $\tilde p_{\pi}$),
which are antisymmetric and symmetric with respect to reflection in  
M$_1$,
respectively.
The precise definition of these orbitals will be given in the next  
section.

Thus we are able to predict the symmetry of each band
at the high symmetry points in the Brillouin zone (BZ).
However, as will be seen later, the order of energy levels of the LDA  
calculation is
incompatible with the experimental spectra. 
Moreover, as it was mentioned already, LDA calculations are unable to describe
the Mott insulating ground states of the undoped cuprates and do not produce
the Cu local moments that are present in these systems. The splitting of the
spectral density due to the 3$d_{x^2-y^2}$ states 
away from the 
Fermi energy due to Coulomb correlations and the resulting reduction in 
Cu-O hybridization 
is expected to
be largely missing in such calculations. However, what, if any, changes there
are from the LDA bands away from $E_F$ is unclear, particularly well above
$T_N$, where the magnetic scattering due to antiferromagnetic spin
fluctuations should be more or less incoherent. Addressing this question is
one of the main goals of the present paper. In the following we develop 
a more sophisticated LDA+U calculation taking into account  
explicitly
the effects of strong correlations. 
As a preliminary step we formulate an effective tight-binding
model which will be fitted both to the LDA+U band structure
calculations and the ARPES VB spectra.

\section{Symmetry analysis and tight-binding model}
The polarization dependent ARPES measurements of VB states along the  
two
high-symmetry directions $\Gamma-(\pi,\pi)$ and $\Gamma-(\pi,0)$  
discriminate
the parity of these states with respect to reflections in the  
corresponding
mirror planes M$_1$ and M$_2$. To make the analysis of the experimental  
data
more straightforward it is helpful to incorporate the symmetry  
properties of
the VB states in our approach from the beginning. This becomes  
especially clear by
constructing an effective tight-binding (TB) model taking into account  
the point-group
symmetry of the VB states. The TB model will be
restricted to the 11 bands of Cu 3$d$ and O 2$p$. Of course, as can be  
seen
in Table 1, there occurs in some cases quite a strong mixing with the  
Cl
subsystem, but in the following we will assume that this mixing is  
taken into
account by the particular values of the TB parameters.

We start with the description of in-plane oxygen orbitals whose  
analysis is
more involved than that for the copper or out-of-plane orbitals. We  
introduce
the annihilation operator of an electron in the two oxygen  
$\pi$-orbitals belonging to
an elementary cell at position $\vec i$ ($\vec i$ is a site of the  
square
lattice) as $p_{i+\beta/2}^{(\alpha)}$, where $(\vec \alpha, \vec  
\beta) =
(\vec x, \vec y)$ or $(\vec y,
\vec x)$ with $\vec x$ and $\vec y$ to be the two orthogonal unit
vectors of the lattice. The $d_{xy}$ orbitals
hybridize with a particular combination of
oxygen orbitals arranged over the plaquette at site $\vec i$:
$p_{\pi i} =\frac{1}{2}
(p_{i+x/2}^{(y)}-p_{i-x/2}^{(y)}+p_{i+y/2}^{(x)}-
p_{i-y/2}^{(x)})$.
This plaquette's $\pi$-orbitals are not orthogonal to each other. The
orthogonalization can be made by introducing first the Fourier
transformation for the original $p_{\pi}$-orbitals
$$
p_{\pi}^{(\alpha)}(q) = \frac{1}{\sqrt{N}} \sum_i
p_{i+\beta/2}^{(\alpha)} e^{-i\vec q (\vec i + \vec \beta /2)}  \; .
$$
At the second
step we define two kinds of canonical Fermi-operators
\begin{eqnarray}
p_{\pi} (q)&=& \lambda_{q}^{-1} i (s_{q,y} p_{\pi}^{(y)} (q) -
                                  s_{q,x} p_{\pi}^{(x)} (q))
\nonumber \\
\tilde p_{\pi} (q) &=& \lambda_{q}^{-1} i (s_{q,x} p_{\pi}^{(y)} (q)  
+
                                  s_{q,y} p_{\pi}^{(x)} (q))
\end{eqnarray}
where $s_{q,\alpha}= \sin (q_{\alpha}/2)$ ($\alpha=x,y$) and  
$\lambda_q =
\sqrt{s_{q,x}^2+s_{q,y}^2}$.
It is easy to see that $p_{\pi}$ and $\tilde p_{\pi}$ are orthogonal
with respect to each other. The definition (1) provides
an equivalent representation for $\pi$-orbitals in terms  of
$p_{\pi} (q)$ and $\tilde p_{\pi} (q)$, instead of the original
$p_{\pi}^{(x)} (q)$ and $p_{\pi}^{(y)} (q)$
operators and takes into
account the point group symmetry of the CuO$_2$ plane. In particular,  
for $q$
along
$\Gamma-(\pi,\pi)$, the $p_{\pi}$-orbital is antisymmetric with respect
to reflections in the mirror plane $M_1$, while the $\tilde
p_{\pi}$-orbital is symmetric (see Fig.\ 4). Along $\Gamma-(\pi,0)$, we find
$p_{\pi}$ to be symmetric and $\tilde p_{\pi}$ to be
antisymmetric with respect to reflection in $M_2$.

Turning now to the oxygen $\sigma$-orbitals we carry out the same  
procedure as
above with the corresponding $p_{i+\alpha/2,\sigma}^{(\alpha)}$  
operators
($\vec \alpha = \vec x, \vec y$).
In this case, introducing the plaquette representation
instead of defining the original $p_{\sigma}^{(\alpha)}$ operators in  
momentum space,
we define a new pair of canonical Fermi-operators $p_{\sigma}$ and  
$\tilde p_{\sigma}$:
$$
p_{\sigma} (q) = \lambda_{q}^{-1} i (s_{q,x} p_{q\sigma}^{(x)} -
                                  s_{q,y} p_{q\sigma}^{(y)})
$$
$$
\tilde p_{\sigma} (q) = \lambda_{q}^{-1} i (s_{q,y} p_{q\sigma}^{(x)}  
+
                                  s_{q,x} p_{q\sigma}^{(y)})
\; .
$$
The notation is chosen in such a way that the $p_{\sigma} (q) (\tilde
p_{\sigma} (q))$-orbitals have the same
symmetry properties with respect to reflections at M$_1$ and M$_2$ as  
the
$p_{\pi} (q)$ or $\tilde p_{\pi} (q)$-orbitals,
respectively.

The definition of the corresponding copper annihilation operators is  
quite
standard and thus we may write down the TB Hamiltonian
\begin{equation}
H_t = \sum_{q \mu \nu s} c_{\mu s}^{\dagger} (q) H_{\mu \nu} (q)
c_{\nu s} (q) \; . 
\label{tb}
\end{equation}
Here, $c_{\mu s}$ is an annihilation operator of either an oxygen $p$ orbital
or a copper $d$ orbital, 
where the indices $\mu$ and $\nu$ denote the 11 different orbitals and  
$s$
denotes the spin.
All orbitals can be classified as to whether they hybridize in-plane or
out-of-plane and there
is no coupling between the two subsystems. The
orbitals involved in the hybridisation in-plane are $p_{\sigma}$,  
$p_{\pi}$,
$\tilde p_{\sigma}$,
$\tilde p_{\pi}$,
$d_{x^2-y^2}$, $d_{xy}$, $d_{3 z^2-r^2}$.
The explicit form of the
TB-Hamiltonian for in-plane orbitals is given in the Appendix.
The in-plane part of the TB-model has 11 parameters: the on-site
energies $\varepsilon_d$ (for $d_{x^2-y^2}$), $\varepsilon_D$ (for  
$d_{xy}$)
and $\varepsilon_{\tilde d}$ (for $d_{3z^2-r^2}$) as well as  
$\varepsilon_p$
(corresponding to $p_{\sigma}$) and $\varepsilon_{\pi}$; the hopping  
matrix
elements $t_{pd}$, $t_{pD}$, $t_{p \tilde d}$,
$t_{pp}$, $t_{\pi\pi}$ and $t_{p\pi}$.
Besides the orbitals hybridising in-plane we have to consider those  
involved
in hybridization out-of-plane:
O 2$p_z$,  Cu 3$d_{xz}$ and Cu 3$d_{yz}$. Restricting ourselves to
nearest
neighbor hopping leads to two 2 $\times$ 2 matrices with on-site  
energies
$\varepsilon_{pz}$ and $\varepsilon_{dz}$ and the hopping matrix  
element
$t_{pdz}$.

In order to analyze the experiment it is important to know the parity  
of the
orbitals with respect to
reflections at the corresponding mirror planes M$_1$ and M$_2$. This  
can also 
be
expressed in terms of group theory since for $k$ vectors along the line
$\Gamma-(\pi,\pi)$ all wave functions can be classified in terms of
irreducible representations of the small group $C_{2v}$. 
\cite{wigner,luehrmann,kettle} The
bands built up from the in-plane orbitals
$d_{xy}$, $d_{3 z^2-r^2}$, $\tilde p_{\sigma}$ and $\tilde p_{\pi}$  
belong to
the representation $A_1$ and are symmetric with respect to reflections  
at
M$_1$, whereas $d_{x^2-y^2}$, $p_{\sigma}$ and  $p_{\pi}$ belong to  
$A_2$ and
are antisymmetric. The same small group $C_{2v}$ also acts along  
$\Gamma -
(\pi,0)$ and the subdivision of the in-plane orbitals is as follows:
$A_1$ (symmetric): $d_{x^2-y^2}$, $d_{3 z^2-r^2}$, $p_{\sigma}$,  
$p_{\pi}$ 
and
$A_2$ (antisymmetric): $d_{xy}$, $\tilde p_{\sigma}$, $\tilde p_{\pi}$.  
The
subdivision along high-symmetry lines is also easily seen in the  
TB-matrix
given in the Appendix. The corresponding small groups at the  
high-symmetry
points $\Gamma$, $(\pi,\pi)$ and $(\pi,0)$ are $D_{4h}$ and $D_{2h}$,
respectively,   
and the
assignment of the different orbitals to the corresponding irreducible
representations is given in Table 2. Of course, the group theoretical
analysis is not only valid for the TB model but also for the LDA bands 
 (Table 1).

The TB-Hamiltonian (\ref{tb}) should be completed by an interaction  
term
\begin{equation}
 H=H_t+H_U
\label{htu}
\end{equation}
which will not be written out explicitly.
This is just a direct extension of the three-band Emery model to the  
case
of the complete set of 11 bands for the CuO$_2$ plane. The interaction
term $H_U$ involves intrasite Hubbard repulsion for different kinds of
copper and oxygen orbitals and appropriate intersite copper-oxygen
repulsions. In order to establish the one-electron parameters entering  
into
Eq.\ (\ref{htu}) one has to keep in mind that these parameters are  
``bare''
ones while the results of the band structure calculations should be
interpreted in terms of a mean-field solution of Eq.\ (\ref{htu}).
\cite{hyb2} To arrive at the bare parameters, one would have to take  
into
account the ground-state (GS)
properties of the CuO$_2$ plane and approximate the Coulomb interaction
terms.

In an undoped cuprate compound as Sr$_2$CuO$_2$Cl$_2$, the GS
of a particular CuO$_2$ plane contains one hole per cell which
is shared between $d_{x^2-y^2}$ and $p_{\sigma}$ orbitals.
Thus a convenient description of the GS is to introduce the deviations
$\langle n^{d}_{s} \rangle_h = 1 - \langle n^{d}_{s} \rangle$ and
$\langle n^{p}_{s} \rangle_h = 1 - \langle n^{p}_{s} \rangle$ from
the full band (Cu 3$d^{10}$ O 2$p^6$) electron occupancy.
A rough estimate is $\langle n^{d}_{s} \rangle_h \approx 0.7$ and
$\langle n^{p}_{s} \rangle_h \approx 0.3$.
Here $\langle n^{p}_{s} \rangle$ means the electron number in the  
$p_{\sigma}$
orbital with spin $s$ (one should note that the occupation of a local  
oxygen 
orbital is only half that number).
Now the mean-field (``screened'') one-electron energies
$\bar \varepsilon_{\mu s}$ read as follows
\begin{eqnarray}
\bar \varepsilon_{ds} &=& \varepsilon_d + U_d
 -U_{d} \langle n^{d}_{\bar s} \rangle_h - 2 U_{pd}
\langle \sum_{s^{\prime}} n^{p}_{s^{\prime}} \rangle_h
\nonumber \\
\bar \varepsilon_{ps} &=& \varepsilon_{p} + U_p
 - 2 U_{pd} \langle \sum_{s^{\prime}} n^{d}_{s^{\prime}} \rangle_h
 - \frac{1}{2} U_p \langle n^{p}_{\bar s} \rangle_h
\nonumber \\
\bar \varepsilon_{D} &=& \varepsilon_{D} + U_d
 - U_{dD} \langle \sum_{s^{\prime}} n^{d}_{s^{\prime}} \rangle_h
 - 2 U_{Dp} \langle \sum_{s^{\prime}} n^{p}_{s^{\prime}} \rangle_h
\nonumber \\
\bar \varepsilon_{\pi} &=& \varepsilon_{p} + U_{\pi}
 - \frac{1}{2} U_{p\pi} \langle \sum_{s^{\prime}}
n^{p}_{s^{\prime}} \rangle_h
 - 2 U_{d \pi} \langle \sum_{s^{\prime}} n^{d}_{s^{\prime}}
\rangle_h
\; ,
\label{e5}
\end{eqnarray}
where $\bar s = -s$. There are also similar expressions for
$\bar\varepsilon_{\tilde{d}}$,
$\bar\varepsilon_{dz}$, $\bar\varepsilon_{pz}$ which we do not specify  
here.

In the paramagnetic LDA band structure where the correlation effects are
treated only in an averaged manner, the screening effect is nearly the
same for all $d$-levels. So, in the LDA approach the effects of strong
correlations due to $U_d$ are missed. An obvious way to adopt these
effects is to treat the ferromagnetic solution by putting, for  
instance,
$\langle n^{d}_{\downarrow} \rangle_h=0$,
and
$\langle n^{d}_{\uparrow} \rangle_h=n^d$.
Then
$\bar \varepsilon_{d \uparrow}=\varepsilon_d + U_d - 2 U_{pd} \ n^p$
($n^p=\sum_s \langle n^{p}_{s} \rangle_h$),
is shifted upwards while $\bar \varepsilon_{d_\downarrow}= \varepsilon_d + U_d
(1-n^d) - 2 U_{pd} n^p$ is shifted equally downwards with respect to the
paramagnetic solution.
Regarding the other $d$-levels, let us assume for the moment the
rough estimate for the intrasite Coulomb parameters $U_{dD} \simeq  
U_d$.
Then one can see that $\bar \varepsilon_D = \varepsilon_D + U_d  
(1-n^d) - 2 U_{pd} \ n^p$,
and the $d_{xy}$ level as well as all the other remaining Cu $d$-levels
are shifted as was the lower $\bar \varepsilon_{d \downarrow}$.
The spin dependence of $\bar \varepsilon_{ps}$ in (\ref{e5})
is much less pronounced than for $\bar \varepsilon_{ds}$ and is  
neglected in the following.

Thus, although being somewhat awkward, the ferromagnetic solution  
provides a
better description of the strong electron correlations, giving a more
reasonable energy position and
occupancy of the different orbitals. Just this approach
is taken by us to carry out the LDA+U calculation. The details of the
procedure and some results of these calculation are presented in the  
next
Section.

\section{LDA+U calculation}
The main effect of a mean-field treatment of the multi-band Hubbard  
model
is a shift of the on-site copper energies against the oxygen ones.
Furthermore, the on-site energy of the Cu 3$d_{x^2-y^2}$ orbital is  
split
into one for spin up $\bar \varepsilon_{d \uparrow}$ (minority spin)  
and
$\bar \varepsilon_{d \downarrow}$ (majority spin). This can also be  
achieved
by an LDA+U calculation \cite{anisimov} including all valence orbitals.

We performed LDA+U calculations for Sr$_2$CuO$_2$Cl$_2$ using a
ferromagnetic splitting. The on-site energy of the unoccupied, spin up
Cu 3$d_{x^2-y^2}^{\uparrow}$ orbital (minority spin)
is shifted by 2 eV upwards and the occupied, spin down
Cu 3$d_{x^2-y^2}^{\downarrow}$ orbital (majority spin),
as well as both spin directions for all the remaining Cu 3$d$ orbitals are  
shifted
by 2 eV downwards. The energy shifts were added at each step of the
self-consistency cycle until the charge-distribution was stable. We
did not try to connect the chosen energy shifts with the model
parameters such as, for instance, $U_d, U_{pd}, U_p$. According to  
(\ref{e5}), the
actual shift depends also on the occupation numbers
$\langle n^{d}_{s} \rangle_h$ and
$\langle n^{p}_{s} \rangle_h$.
Since we did not shift the oxygen levels, our choice
corresponds in fact to the difference between $U_d$ and $U_p$ weighted
with the corresponding occupation numbers.

The results of our LDA+U calculation  are presented in Fig.\ 5 and  
Table 3.
The mainly unoccupied, minority band of $d_{x^2-y^2}$ and $p_{\sigma}$
character can be roughly interpreted as the upper Hubbard band. The
corresponding band for majority spin lies just below the Fermi level  
and
has dominantly oxygen character. Since its spin is opposite to the spin
of the copper hole, there is some justification to interprete that band
as the mean field representation of the Zhang-Rice singlet. But due to  
our
ferromagnetic spin structure it has completely the wrong dispersion  
relation. \cite{yure} 
The bandwidth of both bands is expected to be strongly reduced by
correlation effects in comparison with Fig.\ 5 such that a gap opens.

Next in binding energy we find bands with dominantly oxygen character.
The nonbonding oxygen band  with lowest binding energy at $(\pi,\pi)$  
is
identified to be of pure $p_{\pi}$ character. The
oxygen bands occur at nearly the same energy for both spin
directions. In fact, only the bands with a considerable weight of the
Cu 3$d_{x^2-y^2}$ orbital show a strong splitting between spin up and
spin down.
Therefore we present in Table 3
only the position of minority spin bands and both spin directions for  
bands
with a contribution from the 3$d_{x^2-y^2}$ orbital. \cite{remark1}
The actual value of the energy shifts of the copper bands in our LDA+U
calculation has little influence on the upper oxygen bands, only
their copper character is changed. We have chosen such a shift that the
copper bands are at the lower edge of the valence band, but are not yet
split off the valence band. This is important in order to achieve
good agreement with the experimental results.

Let us now compare the LDA and LDA+U results starting at $(\pi,\pi)$.  
In both
cases (Figs.\ 2 and 5), we
find a group of 5 bands at around 3 eV binding energy, but the order of  
energy
levels is completely different in the two cases. For example, the 
antisymmetric
$p_{\pi}$
band has lowest binding energy of $\approx$ 2.5 eV in the LDA+U
calculation. In Fig.\ 2 (LCAO-LDA), however, all the other 4 bands of  
that
group have lower binding energy than the $p_{\pi}$ level. And also in the FLAPW
calculation (Fig.\ 3) the
$p_{\pi}$ band has 0.3 eV larger binding energy than the valence band edge.
A similar rearrangement of energy levels can be observed at the  
$\Gamma$
point. Due to symmetry reasons there is no hybridization between copper  
and
oxygen bands there. The energy position of the oxygen bands is nearly  
the 
same
for LDA and LDA+U, but the copper bands are shifted.
The in-plane oxygen bands are twofold degenerate and occur
twice in the LDA+U result with binding energies of 2.69 and 5.57 eV,
respectively.

\section{Comparison with experiment}

\subsection{High-symmetry points}

The experimental ARPES spectra at the high symmetry points for
both polarization directions are presented in Fig.\ 6. At the $\Gamma$  
point, 
there are two possible orientations of the sample such that one can  
probe
the symmetry of states with respect to reflections in either M$_1$  
(sample
directed such that the photoelectron momentum is along
$\Gamma-(\pi,\pi)$, Fig.\ 6a), or M$_2$ (sample directed
such that the photoelectron momentum is along $\Gamma-(\pi,0)$, Fig.\  
6b).
The first peak at 2.9 eV binding
energy in the experimental spectra at the $\Gamma$ point with the  
sample
oriented such that the $k$-vector is along $\Gamma-(\pi,\pi)$ (Fig.\ 6a)  
is equally strong for both
polarization directions. This leads us to interprete it as the two pure  
oxygen bands $(p_{\pi}p_{\sigma})$
and $(\tilde p_{\pi} \tilde p_{\sigma})$ which are antisymmetric and
symmetric with respect to reflections at M$_1$, respectively.
\cite{remark2} These bands occur in the LDA+U calculation as the  
two-fold
degenerate in-plane oxygen bands at 2.69 eV binding energy.
According to this interpretation we would expect the same identical  
peak for both
spin directions also at the $\Gamma$ point with the sample oriented  
such that the $k$-vector is along 
$\Gamma-(\pi,0)$ (Fig.\ 6b). As one can see,
Fig.\ 6b deviates only slightly from that expectation. In the LDA  
result,
however, there are 3 copper levels between 2.3 and 3 eV binding energy.  
Since
every copper level has different symmetry properties with respect to  
M$_1$ 
and M$_2$ that would lead to strong differences between both  
polarization directions
which is not observed. Therefore, we assign each experimental
peak with the help of the LDA+U results. Each pure
band is denoted by one orbital only.
For the mixed bands we choose a notation using two orbitals,
where the first one is the dominant one. The experimental peak  
positions are
compared with the LDA+U positions in Table 4.

Let us continue our interpretation of the spectra at the $\Gamma$ point  
with
the peak at 3.9 eV. It is seen with horizontal polarization in Figs.\  
6a and 6b. Therefore, we interprete
it as the out-of-plane oxygen $p_z$ orbital. We observe also
a small contribution of this peak with the ``wrong'' polarization in  
Fig.\
6a which is even larger in Fig.\ 6b. However, there is no band with the
corresponding symmetry in that energy region in our LDA+U calculation.
The large peaks at around 6 eV binding energy in Figs.\ 6a and 6b with  
big
differences between both polarization directions indicate
that there are additional contributions besides the oxygen orbitals  
there.
Due to the low cross section of Cl 3$p$ orbitals, we are only left with  
the
pure copper $d$ orbitals. To simplify the analysis we did not try to  
assign
the Cu 3$d_{3z^2-r^2}$ orbital which mixes strongly with the Cl  
orbitals
and should have reduced intensity. The remaining in-plane copper  
orbitals
change their polarization dependence between Fig.\ 6a and 6b. The
$d_{x^2-y^2}^{\downarrow}$ is antisymmetric with respect to M$_1$ and  
the
$d_{xy}$ is symmetric, but with the sample oriented such that the  
$k$-vector is along 
$\Gamma-(\pi,0)$ this situation is reversed. 
The intensity ratio between horizontal and vertical polarization of
the peak at 5.8 eV is indeed exchanged if we compare Fig.\ 6a and  6b.
The last peak at 6.5 eV occurs for both sample orientations only with
horizontal polarization and is interpreted as the out-of-plane $d_{xz}$  
or
$d_{yz}$ orbital.

Turning now to the spectra at $(\pi,\pi)$ we can only
probe the parity with respect to M$_1$ (Fig.\ 6c). 
The small prepeak at 1.2 eV in the curve with
vertical polarization is ususally interpreted as the Zhang-Rice
singlet. \cite{wells}
The
dominant peak at 2.4 eV binding energy in the spectra with
perpendicular polarization can be identified as the pure $p_{\pi}$
orbital which
has already been discussed in Ref.\ \onlinecite{pothuizen}.
The $p_{\pi}$ band is the only one among the group of 5 bands at around  
3 eV
binding energy in both calculations (LDA or LDA+U, Figs.\ 2 and 5)  
which
has odd symmetry with respect to M$_1$. It has lowest binding energy in the  
experiment 
and
in the LDA+U calculation. That indicates that the LDA+U calculation is  
better
in predicting the correct order of energy levels at high symmetry  
points than
the pure LDA calculation.
At slightly higher binding energy at 2.7 eV we observe a smaller,  
broader peak with horizontal polarization.
According to our calculation it should be comprised of three bands, the
out-of-plane $(p_z d_{(x,y)z})$ bands and the in-plane $(\tilde p_{\pi}
d_{xy})$ band.
The small structure at 3.8 eV binding energy (vertical polarization)  
can
be related to the oxygen $p_{\sigma}$ orbital hybridizing with
$d_{x^2-y^2}$ but having opposite spin ($\uparrow$) than that of the  
copper
hole. The
corresponding band occurs in the LDA+U at 4.94 eV binding energy and  
can
be interpreted as the Zhang-Rice triplet. A similar structure was also
observed in our previous analysis of the polarization dependent  
photoemission
spectra of another undoped model cuprate Ba$_2$Cu$_3$O$_4$Cl$_2$.
\cite{schmelz}

The peaks at around 6 eV binding energy should be assigned to bands
with a dominant copper character. But one may note in Table 4 a  
systematic
deviation between experimental and theoretical peak positions at
$(\pi,\pi)$: the
theoretical binding energies are too large. That is plausible since
it is expected that the copper bands feel the antiferromagnetic  
correlations
much more than the oxygen bands which are decoupled from the copper  
spins.
As a result the copper bands are expected to follow more the AFM BZ  
where
$\Gamma$ and $(\pi,\pi)$ are identical. However, such AFM correlations  
were 
not
considered in our calculation.

At $(\pi,0)$ (Fig.\ 6d) one may observe a prepeak with low intensity
which may be prescribed to the Zhang-Rice
singlet state comprised in our calculation by the hybridization
between the $p_{\sigma}$ orbital and $d_{x^2-y^2}^{\downarrow}$.
The strong peak with horizontal polarization at 2.5 eV is assigned to  
the
out-of-plane $(p_z d_{xz})$ orbital. The peak at 3.8 eV
consists of two orbitals $p_z$ and $p_{\pi}$ which are separated by
only 0.5 eV in the LDA+U calculation. Therefore it is difficult
to use that peak to extract the parameter $t_{\pi\pi}$ from the  
experimental
spectra as it was done in Ref. \onlinecite{pothuizen}. Furthermore, one  
should distinguish between different oxygen hopping matrix elements  
($t_{pp}$, 
$t_{p\pi}$ and $t_{\pi\pi}$)\cite{mattheiss} which was also not done  
there.\cite{pothuizen}

\subsection{Dispersion relations}

The experimental spectra along both high symmetry directions show clear
differences between both polarization directions (Figs.\ 7-10). The  
first
electron removal peak along $\Gamma - (\pi,\pi)$ has minimal binding energy at
$(\pi/2,\pi/2)$ and occurs exclusively with vertical polarization  
(Fig.\
7). That is in complete agreement with the usual interpretation of that  
peak
as the Zhang-Rice singlet. In our mean-field treatment it is built up of the
$d_{x^2-y^2}^{\downarrow}$ and $p_{\sigma}$ orbitals having odd  
symmetry with
respect to M$_1$. The dispersion is well described within the extended 
$t$-$J$ model \cite{yushankhai} and we have included the corresponding  
theoretical
curve in Fig.\ 7 for completeness.
Along $\Gamma - (\pi,0)$ (Figs. 9 and 10) the Zhang-Rice singlet feature is
less   
pronounced 
and according to our symmetry analysis based on a simple mean-field  
treatment we
would expect it only with horizontal polarization. However it is more  
clearly seen in Fig.\ 9
(vertical polarisation) than in Fig.\ 10 (horizontal polarisation). The
explanation 
of that effect deserves obviously a more refined treatment and will be studied
both theoretically and experimentally in the future. 

The peak next in binding energy in Fig.\ 7
was already analyzed as the $p_{\pi}$ orbital and it has a clear  
dispersion
going from $\Gamma$ to $(\pi,\pi)$.
The valence band edge at around 2.5 eV binding energy is different for both 
polarizations along $\Gamma - (\pi,0)$ as well: it has no dispersion for
vertical polarization (Fig.\ 9)   
and is built up of
only one ($\tilde p_{\pi}$) orbital. In contrast to that, we see for 
horizontal polarization (Fig.\ 10) one dispersionless out-of-plane band  
at
3.9 eV and two crossing bands from the out-of-plane orbitals and the  
in-plane
$p_{\pi}$ band.

To analyse this dispersion quantitatively it is more convenient to use  
the TB
model than the LDA+U calculation due to the restricted number of bands  
in the
former. The parameters of the TB model were found as follows. The LDA+U  
results at high
symmetry points (Table 3) were used to obtain a first parameter set.  
For the
fit we have only chosen such energy levels which have no or very small
contribution from other orbitals (Cu 4$s$, O 3$s$, Cl). In such a way  
our
effective TB parameters also contain the influence of hybridization to  
Cl or
$s$ orbitals. Fitting to the pure LDA results (Table 1) gave nearly the  
same
hopping integrals but different on-site energies. The parameters are  
very
similar to those known for La$_2$CuO$_4$. \cite{hybertsen} After  
fitting to
the LDA+U results there remained small differences to the experimental
dispersions even for the peaks with lowest binding energy. These small
discrepancies to the experimental peak positions were
corrected by small changes of the on-site and off-site energies (here,
especially $t_{pdz}$ was increased). The resulting parameter set is  
shown in
Table 5.

In Fig.\ 11 we have collected all the peak positions from Figs.\ 7-10  
together
with the dispersion of the TB bands. We have distinguished between the
results for vertical polarization (Fig.\ 11a) and horizontal  
polarization
(Fig.\ 11b). According to our previous analysis, the peaks in Fig.\ 11a
between $(\pi,0)$ and $\Gamma$ should only be compared with the 3 TB  
bands
stemming from the $\tilde p_{\sigma}$, $\tilde p_{\pi}$ and $d_{xy}$ 
orbitals.
Analogously, between $\Gamma$ and $(\pi,\pi)$ (Fig.\ 11a) we present  
only the
antisymmetric bands from the $p_{\sigma}$, $p_{\pi}$,
$d_{x^2-y^2}^{\downarrow}$ and $d_{x^2-y^2}^{\uparrow}$ orbitals. In  
Fig.\ 
11, we have collected the bands arising from both the  
$d_{x^2-y^2}^{\downarrow}$ or
$d_{x^2-y^2}^{\uparrow}$ orbitals, and have neglected
the band corresponding to the Zhang-Rice singlet since we cannot expect to
obtain its  
correct dispersion in our simple mean-field treatment. The number of  
bands which
contribute to the spectra for horizontal polarization (Fig.\ 11b) is
considerably larger: these include all of the out-of-plane orbitals and
additionally the corresponding symmetric bands (representation $A_1$ of  
$C_{2v}$) of the
in-plane orbitals.

In Fig. 11 we have distinguished between bands with dominant oxygen
character for all $k$ values (solid lines) and those bands which have a  
considerable
coupling to the copper spins (dashed lines). As one may note, there is  
a considerable 
agreement between experimental and theoretical dispersions for the  
oxygen bands with
small binding energy. Furthermore, there is some similarity at the $\Gamma$
point   
besides
the peak with vertical polarization at 3.9 eV binding energy for which we have
no  
explanation. But the copper bands at around 6 eV disperse strongly in  
the TB
calculation whereas they are nearly dispersionless in the experiment.  
We
think that this failure of the theoretical description arises due to  
the
neglect of antiferromagnetic correlations. To avoid misunderstanding we  
should stress that also the oxygen bands of our mean-field calculation  
have a copper
contribution (except some cases at the high-symmetry points), but that  
the
copper contribution is not dominant. We have also shown the calculated  
dispersion relations
of the oxygen bands in Figs.\ 7 - 10 as solid lines in order to guide  
the eye.

\section{Conclusions}

It can be summarized that polarization dependent ARPES at $\Gamma$,
$(\pi,\pi)$ and $(\pi,0)$ and along the two high-symmetry directions  
gives detailed information
about the bands with different parity with respect to
reflections at the mirror planes M$_1$ and M$_2$. The assignment of the  
peaks
can be performed by means of a symmetry analysis of band structure
results. Here we pick out the three major results.

{\bf Rearrangement of energy levels.} Comparing LDA with LDA+U results  
at
high-symmetry points we found that the strong electron correlation  
leads to a
changed order of energy levels, whereby the experimental peak positions  
could be
more accurately assigned with the help of the LDA+U calculation. In  
comparison with an LDA
calculation we found the copper bands shifted to higher binding energy. So, we  
conclude
that the correlation influences not only the band near the Fermi level  
but
leads to a rearrangement of energy levels throughout the whole VB.

{\bf Check of the non-bonding $p_{\pi}$ band.} Polarization dependent 
ARPES measurements provide a sensitive test of the symmetries of the  
excitations
with low binding energy which were already analyzed before. The $p_{\pi}$
orbital   
is
seen at $(\pi,\pi)$ with vertical polarization as a single peak. At
$(\pi,0)$ it is visible with horizontal polarization but overlaps with
out-of-plane orbitals which makes a parameter assignment difficult.
This means that in polarization independent measurement, such as those  
in Ref.\ \onlinecite{pothuizen}, 
the spectral weight assigned to the $p_{\pi}$ peak at $(\pi,\pi)$ will  
have additional
contributions besides the pure $p_{\pi}$ orbital (of roughly one third of the
total intensity as seen in Fig.\ 6c). As a consequence, the experimental
estimate of the spectral weight of the Zhang-Rice singlet part, which was
performed there using the intensity of the $p_{\pi}$ feature as a 
calibration, should be increased by 50 per cent.

{\bf Dispersion relations.} Analyzing the dispersion relations we  
observe a
difference beween the copper bands which
couple strongly to the antiferromagnetic spin structure and thus feel  
the
antiferromagnetic BZ and the nonbonding oxygen bands which are  
decoupled from
the spin system and follow the paramagnetic (or ferromagnetic) BZ. 
To take that effect  
into
account for Sr$_2$CuO$_2$Cl$_2$ we should extend our theory twice.  
First we
should incorporate the
antiferromagnetic order. Then all the bands are defined within the AFM  
BZ. To
obtain in such a scheme the observed difference between $\Gamma$ and  
$(\pi,\pi)$ 
deserves the calculation of matrix elements.

Despite the fact that the experimental order of energy levels can be 
explained
by an LDA+U calculation one should be aware that the agreement between
photoemission and LDA+U cannot be perfect. First of all, the LDA+U 
calculation
cannot reproduce the satellite structure present in the spectra at  
about 14 
eV
binding energy. And second, the LDA+U has the tendency to push the  
copper
levels to too large binding energy. That was visible in our analysis
especially at
$(\pi,\pi)$. The $k$-integrated copper density of states can also
be measured by x-ray photoemission with large photon energy such that  
the
copper cross section dominates that of oxygen. \cite{boeske} It was  
found 
that
the x-ray photoemission spectrum of the valence band of  
Sr$_2$CuO$_2$Cl$_2$
showed the existence of Cu 3$d$ electron removal states over an energy  
range
of some 5-6 eV.
To compare our LDA+U calculation with earlier ones for La$_2$CuO$_4$
\cite{czyzyk}, one should also keep in mind that we had to choose a  
rather
small shift of the copper levels to find agreement with the  
experimental
situation, and we did not choose correlation parameters from a  
constrained
density functional calculation as in Ref.\ \onlinecite{czyzyk}.
It can be expected that the consideration of self-energy corrections as  
was
done recently by calculating the three-body scattering contributions
\cite{igarashi,takahashi} improves the situation and allows one to work  
with
real
correlation parameters instead of fitted ones. Our main goal here was  
the
assignment of peaks and not the determination
of parameters.
To extract parameters from
polarization dependent ARPES measurements there are several  
improvements
necessary both from the experimental and the theoretical side of view.

\vspace*{1cm}

{\bf Acknowlegements}

\vspace*{1cm}

We are grateful to D. Sch\"afer, V. Theresiak and H. Zhang for carrying  
out
the crystal orientation and thank W. H\"oppner and L. Siurakshina for
technical assistance. Furthermore, we thank J. Igarashi and M. Richter  
for useful discussions.
We acknowledge the financial support of the Max-Planck-Gesellschaft and of the
Heisenberg-Landau program. 
This work was supported by the Deutsche  
Forschungsgemeinschaft
(Graduiertenkolleg der TU-Dresden: "Struktur- und Korrelationseffekte  
in Festk\"orpern") the BMBF
(05-605BDA / 05-SF8BD11) and 'der Fonds der chemischen Industrie'.

\newpage
\appendix
\setcounter {section} {2}
\begin{center}
{\Large\bf Appendix}
\end{center}

The TB-matrix for the in-plane orbitals can be found in the form

\vspace*{1cm}

\begin{tabular}{l||l|l|l|l|l|l|l||}
  & $d_{x^2-y^2}$ & $d_{3z^2-r^2}$ & $d_{xy}$ & $p_{\sigma}$ &  
$p_{\pi}$
& $\tilde p_{\sigma}$ & $\tilde p_{\pi}$ \\
\hline \hline
$d_{x^2-y^2}$ & $\varepsilon_d$ & 0 & 0 & $- 2 t_{pd} \ \lambda_q$ & 0  
& 0 & 
0
\\ \hline
$d_{3z^2-r^2}$ & 0 & $\varepsilon_{\tilde d}$ & 0 &
$t_{p \tilde d} \ \eta_q$
& 0 &
$t_{p \tilde d} \ \beta_q$ & 0 \\ \hline
$d_{xy}$ & 0 & 0 & $\varepsilon_{D}$ & 0 & 0 & 0 &
$ t_{\pi D} \ \lambda_q$ \\ \hline
$p_{\sigma}$ &
$- 2 t_{pd} \ \lambda_q$ &
$t_{p \tilde d} \ \eta_q$ & 0 &
$\varepsilon_p-t_{pp} \ \mu_q$ &
$t_{p\pi} \ \alpha_q$ & $t_{pp} \ \nu_q$ & 0 \\ \hline
$p_{\pi}$ & 0 & 0 & 0 & $t_{p\pi} \ \alpha_q$   &
$\varepsilon_{\pi}-t_{\pi\pi} \ \mu_q$ & 0 &
$- t_{\pi\pi} \ \nu_q$ \\ \hline
$\tilde p_{\sigma}$ & 0 &
$t_{p\tilde d} \ \beta_q$ & 0 & $t_{pp} \ \nu_q$ & 0 &
$\varepsilon_p+t_{pp} \ \mu_q$ &
$-t_{p\pi} \ \alpha_q$   \\ \hline
$\tilde p_{\pi}$ & 0 & 0 &
$ t_{\pi D} \ \lambda_q$ & 0 & $-t_{\pi\pi} \ \nu_q$ &
$-t_{p\pi} \ \alpha_q$ &
$\varepsilon_{\pi}+t_{\pi\pi} \ \mu_q$ \\
\hline
\end{tabular}

\vspace*{1cm}
where $s_{q,x}$, $s_{q,y}$ and $\lambda_q$ are defined in the main
text and the other expressions are given by:

\begin{eqnarray*}
\mu_q&=&\frac{8 s_{q,x}^2 s_{q,y}^2}{\lambda_q^2}
\quad , \quad
\nu_q=\frac{4 s_{q,x} s_{q,y} \left( s_{q,x}^2-s_{q,y}^2 \right) }
{\lambda_q^2} \nonumber \\
\eta_q&=&\frac{s_{q,x}^2-s_{q,y}^2}{\lambda_q}
\quad , \quad
\beta_q=\frac{2 s_{q,x} s_{q,y}}{\lambda_q} \nonumber \\
\alpha_q&=&4 \cos \left( \frac{q_x}{2} \right)
\cos \left( \frac{q_y}{2} \right) \; .
\end{eqnarray*}

\newpage
\appendix
\setcounter {section} {2}
\begin{center}
{\Large\bf FIGURES}
\end{center}

FIG.\ 1. The mirror planes of the CuO$_2$ plane. Filled (open) circles
correspond to copper (oxygen) atoms. 

\vspace*{0.6cm}

FIG.\ 2. The LDA-LCAO band structure of Sr$_2$CuO$_2$Cl$_2$. The wave vector is
given in units of $(\pi/a,\pi/a,\pi/c)$. 

\vspace*{0.6cm}

FIG.\ 3. The LDA-FLAPW band structure of Sr$_2$CuO$_2$Cl$_2$. The points in
$k$-space are denoted as Z=$(0,0,\pi/c)$ and X=$(\pi/a,\pi/a,0)$. 

\vspace*{0.6cm}

FIG.\ 4. Scetch of the different oxygen orbitals within one unit cell (filled
circles - copper, open circles - oxygen) for momenta $q \to 0$ along $\Gamma -
(\pi,\pi)$.

\vspace*{0.6cm}

FIG.\ 5. LDA+U band structure: (a) minority spin ($\uparrow$),
(b) majority spin ($\downarrow$).

\vspace*{0.6cm}

FIG.\ 6. Experimental photoemission data at high-symmetry points, with  
the relevant mirror
plane given in brackets:
(a) at the $\Gamma$ point (M$_1$), (b) at the $\Gamma$ point (M$_2$),
(c) at $(\pi,\pi)$ (M$_1$) and (d) $(\pi,0)$ (M$_2$). The assignment of  
peaks is according to
the LDA+U results.
The filled circles and full lines correspond to vertical
polarization, whereas the open circles and broken lines give the  
results for
horizontal polarization.

\vspace*{0.6cm}

FIG.\ 7. Angle resolved photoemission curves along $\Gamma - (\pi,\pi)$
for vertical polarization. Also shown is the antisymmetric TB-band with
dominant oxygen ($p_{\pi}$) contribution (full line) and the dispersion  
of 
the
Zhang-Rice singlet according to Ref. \onlinecite{yushankhai} (dotted line).

\vspace*{0.6cm}

FIG.\ 8. Angle resolved photoemission curves along $\Gamma - (\pi,\pi)$
for horizontal polarization together with the calculated oxygen  
out-of-plane TB-bands
and the in-plane bands having even symmetry (calculated bands are shown  
as solid lines).

\vspace*{0.6cm}

FIG.\ 9. Angle resolved photoemission curves along $\Gamma - (\pi,0)$
for vertical polarization with the calculated antisymmetric oxygen  
TB-band (shown as a solid line).

\vspace*{0.6cm}

FIG.\ 10. Angle resolved photoemission curves along $\Gamma - (\pi,0)$
for horizontal polarization with the calculated oxygen out-of-plane  
TB-bands and the
in-plane bands having even symmetry (calculated bands are shown as  
solid lines).

\vspace*{0.6cm}

FIG.\ 11. Position of the main experimental peaks together with the  
TB-bands
of the corresponding symmetry along $(\pi,0) - (0,0) - (\pi,\pi)$: (a)
antisymmetric bands and experimental data
for vertical polarization, (b) out-of-plane and symmetric bands  
together
with experimental data for horizontal polarization.
Full lines denote the TB bands with dominantly oxygen character,  
whereas the
dashed lines correspond to bands with a considerable mixing to the  
copper
system.

\newpage

\begin{center}
{\Large\bf TABLES}
\end{center}

TABLE 1: LDA data at high symmetry points showing the weights of the  
different orbital groups
contributing to each band. Also given are the different
reflection symmetries with respect to M$_1$ and M$_2$, respectively
(antisymmetric (A), symmetric (S) and out-of-plane bands (o)).

\vspace{0.3cm}

\begin{center}
{\Large\bf $\Gamma$}
\end{center}

\vspace{0.3cm}
\begin{center}
\begin{tabular}{|c|c|c|c|c|c|c|c|c|c|c|c|c|c|c|}
\hline
No.&
E/eV&
p$_z$&
p$_\sigma$&
p$_\pi$&
d$_{3z^2-r^2}$&
d$_{(x,y)z}$&
d$_{xy}$&
d$_{x^2-y^2}$&
O$_s$&
Cu$_s$&
$\sum$Cl&
not.&
M$_1$&
M$_2$\\
\hline
\hline
1&
 -1.64&
0&
0&
0&
0&
0&
0&
.903&
.097&
0&
0&
d$_{x^2-y^2}$&
A&
S\\
\hline
2&
 -2.28&
0&
0&
0&
.817&
0&
0&
0&
.015&
.005&
.163&
d$_{3z^2-r^2}$&
S&
S\\
\hline
3,4&
 -2.34&
0&
.456&
.530&
0&
0&
0&
0&
0&
0&
.014&
(p$_\pi$p$_\sigma$)&
A&
S\\
&&&&&&&&&&&&(\~{p}$_\pi$\~{p}$_\sigma$)&
S&
A\\
\hline
5&
 -2.72&
0&
0&
0&
0&
0&
1.00&
0&
0&
0&
0&
d$_{xy}$&
S&
A\\
\hline
6,7&
 -2.96&
0&
0&
0&
0&
.984&
0&
0&
0&
0&
.016&
d$_{(x,y)z}$&
o&
o\\
\hline
8,9&
 -3.46&
1.00&
0&
0&
0&
0&
0&
0&
0&
0&
0&
p$_{z}$&
o&
o\\
&
(-3.19)&
(.526)&
&
&
&
&
&
&
&
&
(.474)&&&
\\
\hline
10,11&
 -5.14&
0&
.495&
.495&
0&
0&
0&
0&
0&
0&
.010&
(p$_\sigma$p$_\pi$)&
A&
S\\
&&&&&&&&&&&&(\~{p}$_\sigma$\~{p}$_\pi$)&
S&
A\\
\hline
\end{tabular}
\end{center}
\vspace{0.3cm}

\begin{center}
{\Large\bf ($\pi,\pi$)}
\end{center}

\vspace{0.3cm}
\begin{center}
\begin{tabular}{|c|c|c|c|c|c|c|c|c|c|c|c|c|c|}
\hline
No.&
E/eV&
p$_z$&
p$_\sigma$&
p$_\pi$&
d$_{3z^2-r^2}$&
d$_{(x,y)z}$&
d$_{xy}$&
d$_{x^2-y^2}$&
Cu$_s$&
O$_s$&
$\sum$Cl&
not.&
M$_1$\\
\hline
\hline
1&
2.32&
0&
.554&
0&
0&
0&
0&
.446&
0&
0&
0&
(d$_{x^2-y^2}$p$_\sigma$)&
A\\
\hline
2&
 -1.33&
0&
0&
.196&
.006&
0&
.792&
0&
0&
0&
.006&
(d$_{xy}$\~{p}$_\pi$)&
S\\
\hline
3,4&
 -1.58&
.563&
0&
0&
0&
.437&
0&
0&
0&
0&
0&
(d$_{(x,y)z}$p$_{z}$)&
o\\
\hline
5&
 -1.87&
0&
.038&
0&
.637&
0&
.009&
0&
.055&
0&
.261&
(d$_{3z^2-r^2}$\~{p}$_\sigma$)&
S\\
\hline
6&
 -2.12&
0&
0&
1.00&
0&
0&
0&
0&
0&
0&
0&
p$_\pi$&
A\\
\hline
7,8&
 -4.56&
.641&
0&
0&
0&
.268&
0&
0&
0&
0&
.091&
(p$_{z}$d$_{(x,y)z}$)&
o\\
\hline
9&
 -5.21&
0&
.424&
0&
0&
0&
0&
.576&
0&
0&
0&
(p$_\sigma$d$_{x^2-y^2}$)&
A\\
\hline
10&
 -6.15&
0&
.003&
.702&
.001&
0&
.291&
0&
0&
0&
.003&
(\~{p}$_\pi$d$_{xy}$)&
S\\
\hline
11&
 -7.23&
0&
.495&
0&
.018&
0&
0&
0&
.294&
0&
.193&
(\~{p}$_\sigma$d$_{3z^2-r^2}$)&
S\\
\hline
\end{tabular}
\end{center}
\vspace{0.3cm}

\begin{center}
{\Large\bf ($\pi,0$)}
\end{center}

\vspace{0.3cm}
\begin{center}
\begin{tabular}{|c|c|c|c|c|c|c|c|c|c|c|c|c|c|c|}
\hline
No.&
E/eV&
p$_z$&
p$_\sigma$&
p$_\pi$&
d$_{3z^2-r^2}$&
d$_{yz}$&
d$_{xz}$&
d$_{xy}$&
d$_{x^2-y^2}$&
Cu$_s$&
O$_s$&
$\sum$Cl&
not.&
M$_2$\\
\hline
\hline
1&
 -.40&
0&
.128&
0&
.015&
0&
0&
0&
.599&
.105&
.105&
.049&
(d$_{x^2-y^2}$p$_\sigma$)&
S\\
\hline
2&
 -1.42&
0&
0&
.335&
0&
0&
0&
.665&
0&
0&
0&
0&
(d$_{xy}$\~{p}$_\pi$)&
A\\
\hline
3&
 -1.63&
.395&
0&
0&
0&
0&
.601&
0&
0&
0&
0&
.004&
(d$_{xz}$p$_{z}$)&
o\\
\hline
4&
 -2.12&
0&
.002&
.001&
.655&
0&
0&
0&
.096&
.007&
.019&
.220&
(d$_{3z^2-r^2}$d$_{x^2-y^2}$)&
S\\
\hline
5&
 -2.87&
0&
0&
0&
0&
.880&
0&
0&
0&
0&
0&
.120&
d$_{yz}$&
o\\
\hline
6&
 -3.29&
.594&
0&
0&
0&
0&
.019&
0&
0&
0&
0&
.387&
p$_{z}$&
o\\
\hline
7&
3.58&
0&
.532&
0&
0&
0&
0&
0&
0&
0&
0&
.468&
\~{p}$_\sigma$&
A\\
\hline
8&
 -3.96&
0&
0&
.935&
0&
0&
0&
0&
0&
0&
.046&
.019&
p$_\pi$&
S\\
\hline
9&
 -4.13&
.403&
0&
0&
0&
0&
.270&
0&
0&
0&
0&
.327&
(p$_{z}$d$_{xz}$)&
o\\
\hline
10&
 -4.62&
0&
.057&
.475&
0&
0&
0&
.348&
0&
0&
0&
.120&
(\~{p}$_\pi$d$_{xy}$)&
A\\
\hline
11&
 -5.74&
0&
.268&
.004&
.079&
0&
0&
0&
.149&
.032&
.009&
.459&
(p$_\sigma$d$_{x^2-y^2}$)&
S\\
\hline
\end{tabular}
\end{center}


\newpage


TABLE 2: Assignment of the orbitals to irreducible
representations of the corresponding small groups at high symmetry  
points:
a) $\Gamma$ (group $D_{4h}$), b) $(\pi,\pi)$ ($D_{4h}$) and
c) $(\pi,0)$ ($D_{2h}$). The notations in parantheses are according to
Luehrmann \cite{luehrmann} (see also Ref. \onlinecite{mattheiss}).
Also given are the characters with respect to
reflections at M$_1$ or M$_2$, respectively, whereby + and  -   
correspond to the S and A 
given in Table 1. The orbital $p_z^{(1)}$ means $p_z$ orbitals at  
positions $i \pm x/2$, and $p_z^{(2)}$ at positions $i \pm
y/2$.

\vspace*{1cm}

\begin{center}
{\Large\bf (a) \ $\Gamma$}
\end{center}

\vspace*{1cm}

\begin{center}
\begin{tabular}{|l|ll|c|c|}\hline
\multicolumn{1}{|c|}{orbitals} &
\multicolumn{2}{c|}{repr.} &
\multicolumn{1}{c|}{M$_1$} &
\multicolumn{1}{c|}{M$_2$} \\ \hline
$p_{\sigma}$, $\tilde p_{\sigma}$ & $E_u^{(1)}$ &(5$^-$) & 0 & 0 \\ \hline
$p_{\pi}$, $\tilde p_{\pi}$ & $E_u^{(2)}$ & (5$^-$) & 0 & 0 \\ \hline
$(p_z^{(1)} + p_z^{(2)})/\sqrt{2}$ & $A_{2u}$ & (2$^-$) & + & + \\ \hline
$(p_z^{(1)} - p_z^{(2)})/\sqrt{2}$ & $B_{2u}$ & (4$^-$) & $-$ & + \\ \hline
$d_{x^2-y^2}$   & $B_{1g}$ & (3$^+$) & $-$ & + \\ \hline
$d_{xy}$   & $B_{2g}$ & (4$^+$) & + & $-$ \\ \hline
$d_{(x,y)z}$   & $E_{g}$ & (5$^+$) & 0 & 0 \\ \hline
$d_{3z^2-r^2}$   & $A_{1g}$ & (1$^+$) & + & + \\ \hline
\end{tabular}
\end{center}

\vspace*{1cm}

\begin{center}
{\Large\bf (b) \ ($\pi,\pi$)}
\end{center}

\vspace*{1cm}

\begin{center}
\begin{tabular}{|l|ll|c|}\hline
\multicolumn{1}{|c|}{orbitals} &
\multicolumn{2}{c|}{repr.} &
\multicolumn{1}{c|}{M$_1$} \\ \hline
$d_{3z^2-r^2}$,  $\tilde p_{\sigma}$ & $A_{1g}$& (1$^+$) & +  \\ \hline
$p_{\pi}$ & $A_{2g}$& (2$^+$) & $-$ \\ \hline
$d_{x^2-y^2}$, $p_{\sigma}$ & $B_{1g}$ &(3$^+$) & $-$ \\ \hline
$d_{xy}$, $\tilde p_{\pi}$ & $B_{2g}$ &(4$^+$) & + \\ \hline
$d_{(x,y)z}$, $p_z^{(1,2)}$ & $E_{g}$ &(5$^+$) & 0 \\
\hline
\end{tabular}
\end{center}

\vspace*{1cm}

\begin{center}
{\Large\bf (c) \ ($\pi,0$)}
\end{center}

\vspace*{1cm}

\begin{center}
\begin{tabular}{|l|ll|c|}\hline
\multicolumn{1}{|c|}{orbitals} &
\multicolumn{2}{c|}{repr.} &
\multicolumn{1}{c|}{M$_1$} \\ \hline
$d_{x^2-y^2}$, $d_{3z^2-r^2}$,  $p_{\sigma}$ & $A_{g}$& (1$^+$) & +  \\ \hline
$d_{xy}$, $\tilde p_{\pi}$ & $B_{1g}$& (2$^+$) &$-$ \\ \hline
$\tilde p_{\sigma}$ & $B_{2u}$& (3$^-$) & $-$ \\ \hline
$p_{\pi}$ & $B_{3u}$ &(4$^-$) & + \\ \hline
$d_{xz}$, $p_z^{(1)}$ & $B_{2g}$& (3$^+$) & + \\ \hline
$p_z^{(2)}$ & $B_{1u}$& (2$^-$) & + \\  \hline
$d_{yz}$ & $B_{3g}$& (4$^+$) & + \\ \hline
\end{tabular}
\end{center}


\newpage

TABLE 3: The LDA+U data at the high symmetry points. The bands noted by  
a star
correspond to majority spin ($\downarrow$), whereas all the other data  
are
given for minority spin ($\uparrow$). The column "not." gives the  
notation used to describe the 
bands.

\begin{center}
{\Large\bf $\Gamma$}

\vspace{0.2cm}
\begin{tabular}{|c|c|c|c|c|c|c|c|c|c|c|c|c|c|c|c|}
\hline
No.&
E/eV&
p$_{z}$&
p$_\sigma$&
p$_\pi$&
d$_{3z^2-r^2}$&
d$_{yz}$&
d$_{xz}$&
d$_{xy}$&
d$_{x^2-y^2}$&
O$_s$&
Cu$_s$&
\( \sum  \)Cl&
not.&
M$_1$&
M$_2$\\
\hline
\hline
1&
 -.30&
0&
0&
0&
0&
0&
0&
0&
.909&
.091&
0&
0&
d$^{\uparrow}_{x^2-y^2}$&
A&
S\\
\hline
2,3&
 -2.69&
0&
.439&
.543&
0&
0&
0&
0&
0&
0&
0&
.018&
(\~{p}$_\pi$\~{p}$_\sigma$)&
S&
A\\
&&&&&&&&&&&&&
(p$_\pi$p$_\sigma$)&
A&
S\\
\hline
4,5&
 -3.83&
1.00&
0&
0&
0&
0&
0&
0&
0&
0&
0&
0&
p$_{z}$&
o&
o\\
&
(-3.69)&
(.567)&
&
&
&
&
&
&
&
&
&
(.433)&
&
&
\\
\hline
6&
 -4.58&
0&
0&
0&
.640&
0&
0&
0&
0&
.031&
0&
.329&
d$_{3z^2-r^2}$&
S&
S\\
\hline
7$^\ast$&
 -4.92&
0&
0&
0&
0&
0&
0&
0&
.893&
.107&
0&
0&
d$^{\downarrow}_{x^2-y^2}$&
A&
S\\
\hline
8&
 -5.40&
0&
0&
0&
0&
0&
0&
1.00&
0&
0&
0&
0&
d$_{xy}$&
S&
A\\
\hline
9,10&
 -5.57&
0&
.499&
.463&
0&
0&
0&
0&
0&
0&
0&
.038&
(\~{p}$_\sigma$\~{p}$_\pi$)&
S&
A\\
&&&&&&&&&&&&&
(p$_\pi$p$_\sigma$)&
A&
S\\
\hline
11,12&
 -5.88&
0&
0&
0&
0&
.803&
(.803)&
0&
0&
0&
0&
.197&
d$_{y(x)z}$&
o&
o\\
\hline
\end{tabular}
\vspace{0.2cm}

{\Large\bf ($\pi,\pi$)}

\vspace{0.2cm}
\begin{tabular}{|c|c|c|c|c|c|c|c|c|c|c|c|c|c|c|}
\hline
No.&
E/eV&
p\( _{z} \)&
p\( _{\sigma } \)&
p\( _{\pi } \)&
d\( _{3z^{2}-r^{2}} \)&
d\( _{yz} \)&
d\( _{xz} \)&
d\( _{xy} \)&
d\( _{x^{2}-y^{2}} \)&
O\( _{s} \)&
Cu\( _{s} \)&
\( \sum  \)Cl&
not.&
M\( _{2} \)\\
\hline
1&
3.12&
0&
.467&
0&
0&
0&
0&
0&
.533&
0&
0&
0&
(d$^\uparrow_{x^2-y^2}$p$_\sigma$)&
A\\
\hline
2$^\ast$&
.65&
0&
.704&
0&
0&
0&
0&
0&
.296&
0&
0&
0&
(p$_\sigma$d$^\downarrow_{x^2-y^2}$)&
A\\
\hline
3&
 -2.43&
0&
0&
1.00&
0&
0&
0&
0&
0&
0&
0&
0&
p\( _{\pi } \)&
A\\
\hline
4,5&
 -2.98&
.711&
0&
0&
0&
.143&
.143&
0&
0&
0&
0&
.002&
(p\( _{z} \)d\( _{x(y)z} \))&
o\\
&(-2.97)&&&&&&&&&&&(.006)&&\\
\hline
6&
 -3.35&
0&
.009&
.345&
.030&
0&
0&
.564&
0&
0&
.004&
.048&
(\~{p}\( _{\pi } \)d\( _{xy} \))&
S\\
\hline
7&
 -3.66&
0&
.096&
.020&
.356&
0&
0&
.053&
0&
0&
.051&
.424&
(d\( _{3z^{2}-r^{2}} \)\~p\( _{\sigma } \))&
S\\
\hline
8&
 -4.94&
0&
.561&
0&
0&
0&
0&
0&
.439&
0&
0&
0&
(p\( _{\sigma } \)d$^{\uparrow}_{x^2-y^2}$)&
A\\
\hline
9,10&
 -6.62&
.155&
0&
0&
0&
.341&
.341&
0&
0&
0&
0&
.123&
(d\( _{x(y)z} \)p\( _{z} \))&
o\\
&(-6.06)&(.093)&&&&(.275)&(.275)&&&&&(.357)&&\\
\hline
11&
 -7.20&
0&
.003&
.392&
.010&
0&
0&
.591&
0&
0&
0&
.004&
(d\( _{xy} \)\~p\( _{\pi } \))&
S\\
\hline
12$^\ast$&
 -7.28&
0&
.207&
0&
0&
0&
0&
0&
.793&
0&
0&
0&
d$^{\downarrow}_{x^2-y^2}$p$_\sigma$&
A\\
\hline
13&
 -7.86&
0&
.478&
.001&
.073&
0&
0&
0&
0&
0&
.280&
.168&
(\~p\( _{\sigma } \)d\( _{3z^{2}-r^{2}} \))&
S\\
\hline
\end{tabular}
\vspace{0.2cm}

{\Large\bf ($\pi,0$)}

\vspace{0.2cm}
\begin{tabular}{|c|c|c|c|c|c|c|c|c|c|c|c|c|c|c|}
\hline
No.&
E/eV&
p$_{z}$&
p$_\sigma$&
p$_\pi$&
d$_{3z^2-r^2}$&
d$_{xz}$&
d$_{yz}$&
d$_{xy}$&
d$_{x^2-y^2}$&
O$_s$&
Cu$_s$&
\( \sum  \)Cl&
not.&
M\( _{2} \)\\
\hline
\hline
1&
.47&
0&
.053&
0&
0&
0&
0&
0&
.696&
.124&
.106&
.021&
(d$^{\uparrow}_{x^2-y^2}$p$_\sigma$)&
S\\
\hline
2$^\ast$&
 -2.40&
0&
.273&
0&
.039&
0&
0&
0&
.323&
.083&
.119&
.163&
(p$_\sigma$d$^{\downarrow}_{x^2-y^2}$)&
S\\
\hline
3&
 -2.94&
0&
0&
.659&
0&
0&
0&
.341&
0&
0&
0&
0&
(\~{p}$_\pi$d$_{xy}$)&
A\\
\hline
4&
 -2.96&
.704&
0&
0&
0&
.245&
0&
0&
0&
0&
0&
.051&
(p$_{z}$d$_{xz}$)&
o\\
\hline
5&
 -3.76&
0&
.086&
.003&
.344&
0&
0&
0&
.057&
.006&
.017&
.487&
(d$_{3z^2-r^2}$p$_\sigma$)&
S\\
\hline
6&
 -3.79&
.705&
0&
0&
0&
.013&
0&
0&
0&
0&
0&
.282&
p$_z$&
o\\
\hline
7&
 -4.11&
0&
.635&
0&
0&
0&
0&
0&
0&
0&
0&
.365&
\~{p}$_\sigma$&
A\\
\hline
8&
 -4.32&
0&
0&
.933&
.001&
0&
0&
0&
0&
.049&
0&
.017&
p$_\pi$&
S\\
\hline
9&
 -5.93&
0&
0&
0&
0&
0&
.859&
0&
0&
0&
0&
.141&
d$_{yz}$&
o\\
\hline
10&
 -6.17&
0&
0&
.199&
0&
0&
0&
.801&
0&
0&
0&
0&
(d$_{xy}$\~{p}$_\pi$)&
A\\
\hline
11&
 -6.37&
.128&
0&
0&
0&
.384&
0&
0&
0&
0&
0&
.488&
(d$_{xz}$p$_{z}$)&
o\\
\hline
12&
 -6.44&
0&
.390&
0&
.409&
0&
0&
0&
.104&
.035&
.062&
0&
(p$_\sigma$d$_{3z^2-r^2}$)&
S\\
\hline
13$^\ast$&
 -7.49&
0&
.298&
0&
.060&
0&
0&
0&
.520&
.004&
.076&
.042&
(d$^{\downarrow}_{x^2-y^2}$p$_\sigma$)&
S\\
\hline
\end{tabular}
\end{center}

\newpage

TABLE 4: Comparison of experimental peak positions (in eV)
with the LDA+U results at the high-symmetry points (ZRS and ZRT mean the
Zhang-Rice singlet or triplet, respectively).

\vspace*{0.6cm}

\begin{center}
{\Large\bf $\Gamma$}


\vspace*{0.3cm}


\begin{tabular}{||l|l|l||}
\hline
Orbital     &      LDA+U &  Exp.\  \\
\hline
\hline
$ (p_{\pi} p_{\sigma})$ &  -2.69   &  -2.9   \\
$ (\tilde p_{\pi} \tilde p_{\sigma})$ &   &        \\
\hline
$ p_z $                 &  -3.83   &   -3.9   \\
\hline
$ d_{x^2-y^2}^{\downarrow}$  &  -4.92  &  -5.8  \\
\cline{1-2}
$ d_{xy}$               &  -5.40   &      \\
\cline{1-2}
$ (p_{\sigma} p_{\pi})$ &  -5.57   &     \\
$ (\tilde p_{\sigma}   \tilde p_{\pi})$ &   &        \\
\hline
$d_{(x,y)z}$     &  -5.87   &   -6.5  \\
\hline
\end{tabular}

\vspace*{0.3cm}

{\Large\bf $(\pi,\pi)$}


\vspace*{0.3cm}


\begin{tabular}{||l|l|l||}
\hline
Orbital     &      LDA+U &  Exp.\  \\
\hline
\hline
$ (p_{\sigma} d_{x^2-y^2}^{\downarrow})$ (ZRS)  &  \  0.65   &  -1.2    
\\
\hline
$ p_{\pi} $                 &  -2.43   &   -2.4   \\
\hline
$ (p_z d_{(x,y)z})$  &  -2.98  &  -2.7  \\
\cline{1-2}
$ (\tilde p_{\pi} d_{xy})$               &  -3.35   &     \\
\hline
$ (p_{\sigma} d_{x^2-y^2}^{\uparrow})$ (ZRT)  &  -4.94   &  -3.8   \\
\hline
$(d_{(x,y)z} p_z)$     &  -6.62   &   -5.8  \\
\cline{1-2}
$(d_{xy} \tilde p_{\pi})$     &  -7.20   &     \\
\hline
$ (d_{x^2-y^2}^{\downarrow} p_{\sigma}) $ &  -7.28   &  -6.0   \\
\hline
\end{tabular}

\vspace*{0.3cm}

{\Large\bf $(\pi,0)$}


\vspace*{0.3cm}


\begin{tabular}{||l|l|l||}
\hline
Orbital     &      LDA+U &  Exp.\  \\
\hline
\hline
$ (p_{\sigma} d_{x^2-y^2}^{\downarrow})$ (ZRS)  &  -2.40   &  -1.1   \\
\hline
$ (p_z d_{xz})$  &  -2.96  &  -2.5  \\
\hline
$ (\tilde p_{\pi} d_{xy})$                 &  -2.94   &   -2.7   \\
\hline
$ p_z $  &  -3.79  &  -3.8  \\
\cline{1-2}
$ p_{\pi} $  &  -4.32  &    \\
\hline
$ \tilde p_{\sigma}$ &  -4.11  &  -3.8          \\
\hline
$ (d_{xy} \tilde p_{\pi} )$ & -6.17  &  -5.6       \\
\hline
$ d_{yz}$         &    -5.93   &  -6.6            \\
\cline{1-2}
$ (d_{xz} p_z )$ & -6.37  &        \\
\hline
\end{tabular}
\end{center}

\newpage

TABLE 5: TB-parameters obtained by fitting the LDA+U band structure and  
the VB
photoemission spectra. The off-site energies in
parentheses are the values from a fit only to the theoretical  
band structure in
the cases where experimental corrections were approppriate.


\vspace*{1cm}

\begin{center}
\begin{tabular}{|c|c|c|c|c|c|c|c|}
\hline
$\overline{\varepsilon}^{\uparrow}_{d}$
&$\overline{\varepsilon}^{\downarrow}_{d}$
&$\overline{\varepsilon}_{\tilde{d}}$
&$\overline{\varepsilon}_{D}$
&$\overline{\varepsilon}_{d_z}$
&$\overline{\varepsilon}_{\pi}$
&$\overline{\varepsilon}_{p_z}$
&$\overline{\varepsilon}_{p}$
\\
\hline
\hline
2.00&
 -4.90&
 -4.78&
 -5.22&
 -6.40&
 -3.88&
 -3.86&
 -4.59\\[.3cm]\hline
t$_{pd}$
&t$_{p\tilde d}$
&t$_{pp}$
&t$_{p\pi}$
&t$_{\pi\pi}$
&t$_{\pi D}$
&t$_{pdz}$ & \\
\hline
\hline
1.33&
0.77&
0.71&
0.34&
0.37&
0.84&
1.15&\\
&&&&(0.32)&(0.77)&(0.77)&\\\hline
\end{tabular}
\end{center}
\vspace{0.3cm}


\begin{thebibliography}{30}
\bibitem{remarkmp}
In fact there is a second mirror plane parallel to M$_2$ but going only
through oxygen position. From the point of view of group theory it is
identical to M$_2$ in the paramagnetic case.
\bibitem{remark1}
It should be noted that even the pure oxygen bands have a small energy
difference of 0.17
eV between both spin directions which is the same for all bands. The  
reason
for that difference is the overlap between spin up and spin down
at the Fermi energy. The position of spin down
bands in Table 3 is corrected by this 0.17 eV.
\bibitem{remark2}
In fact, the oxygen in-plane wave functions have to be defined by the  
limit
$k_x=k_y=k \to 0$ to have a defined parity with respect to M$_1$.  
In the 
limit
$k_x= k \to 0$, $k_y=0$ they have a defined parity with respect to  
M$_2$. That
arbitrariness can be explained since they build the two-dimensional
representation $E_u$ at the $\Gamma$ point.
\end{thebibliography}
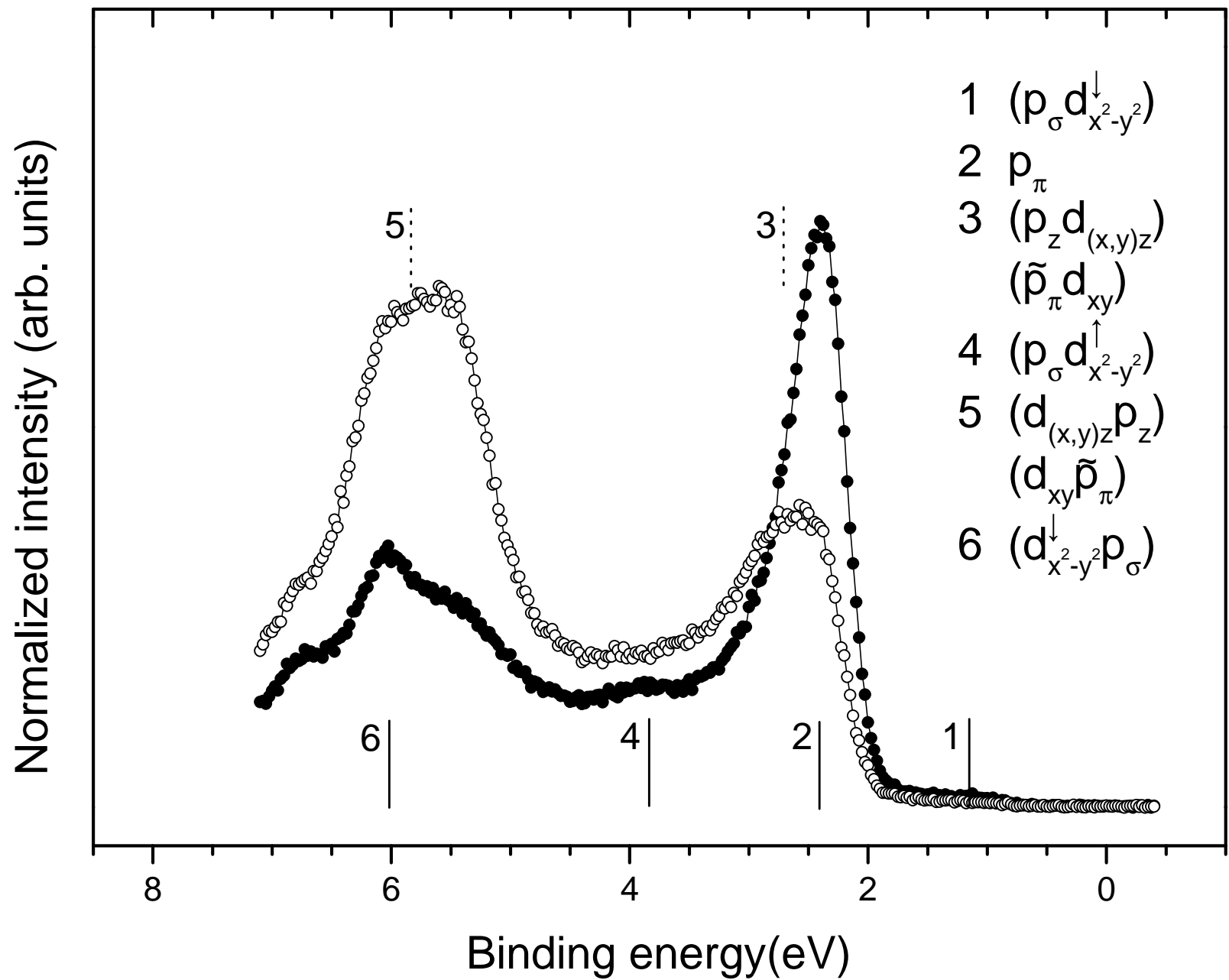
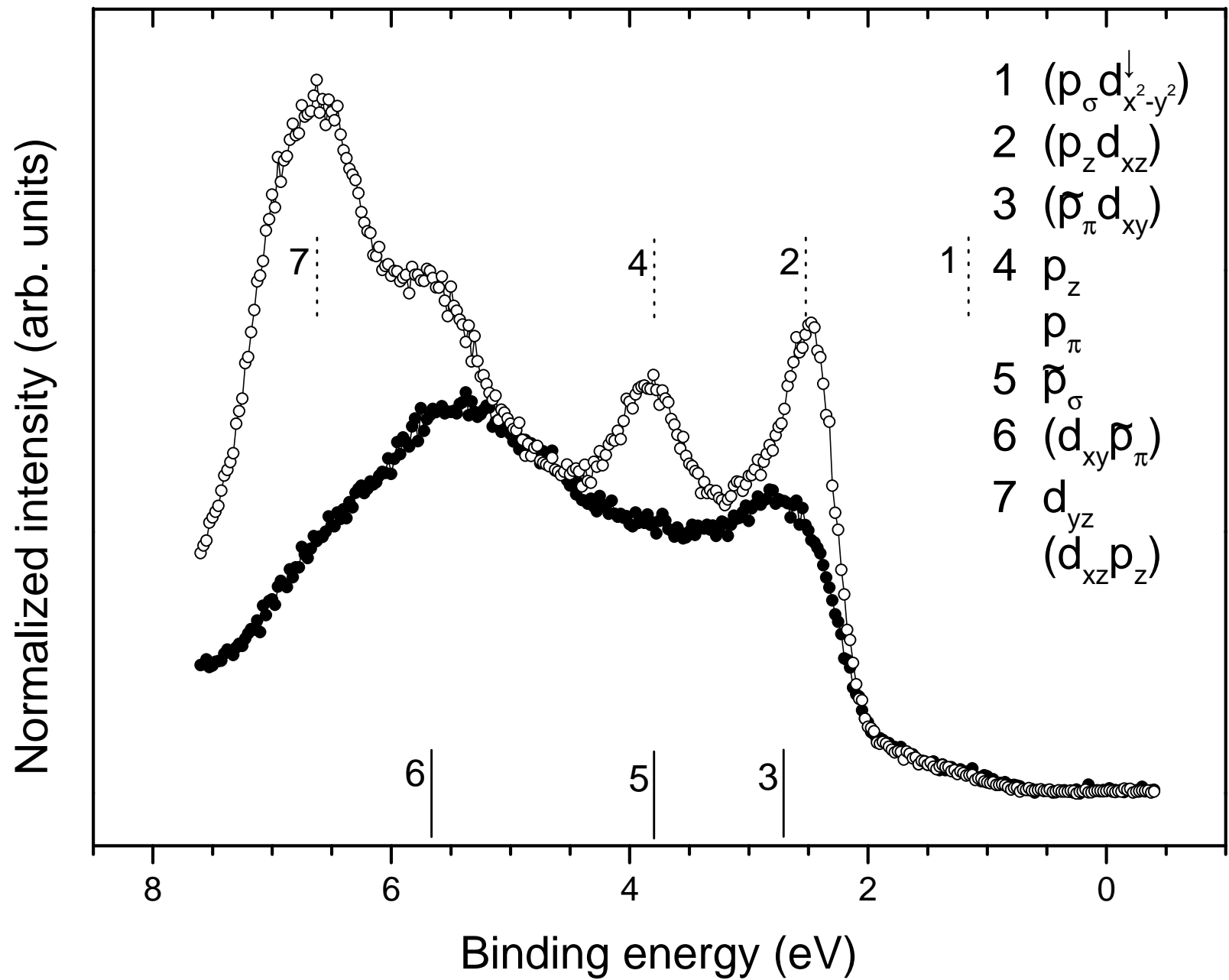
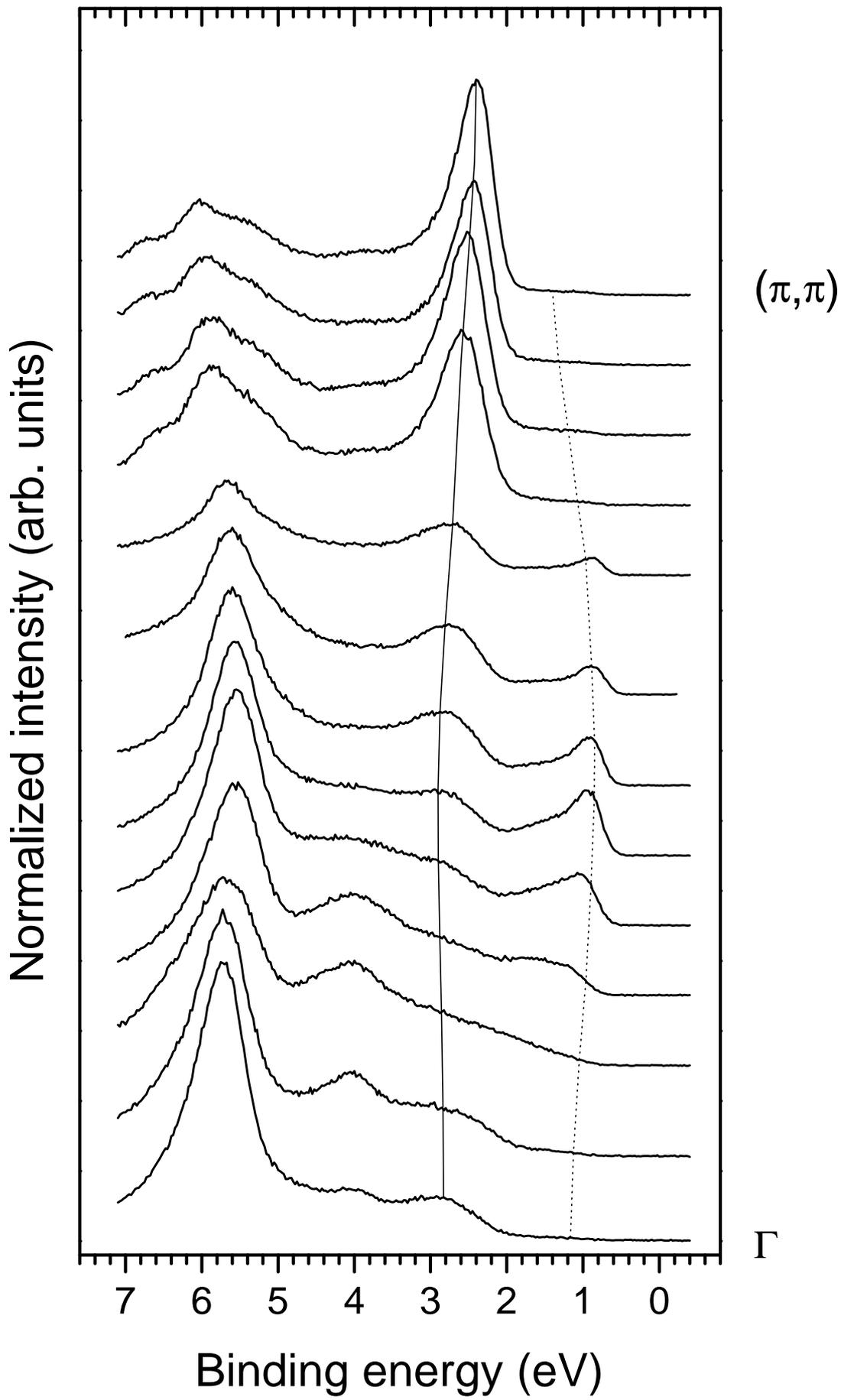
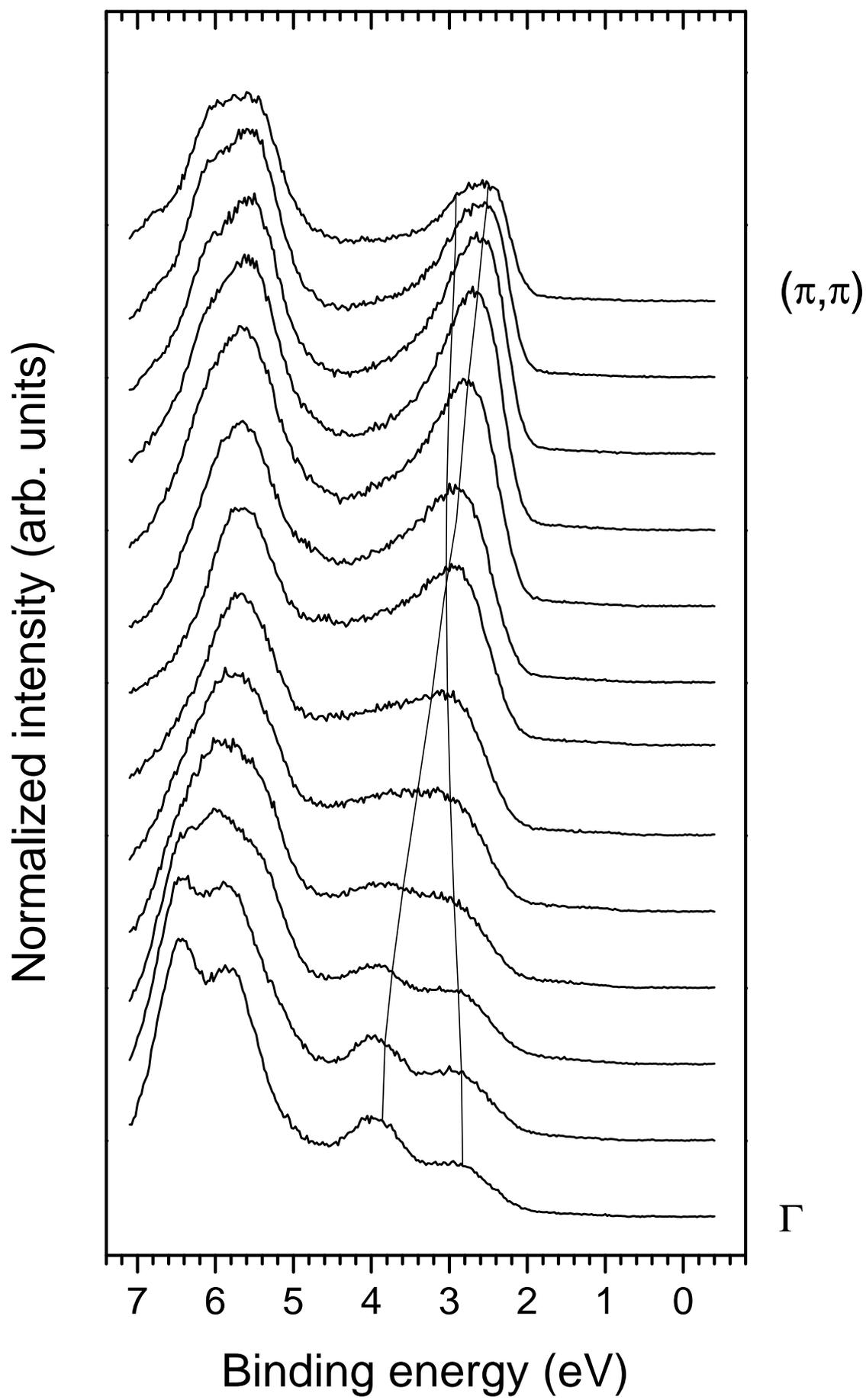
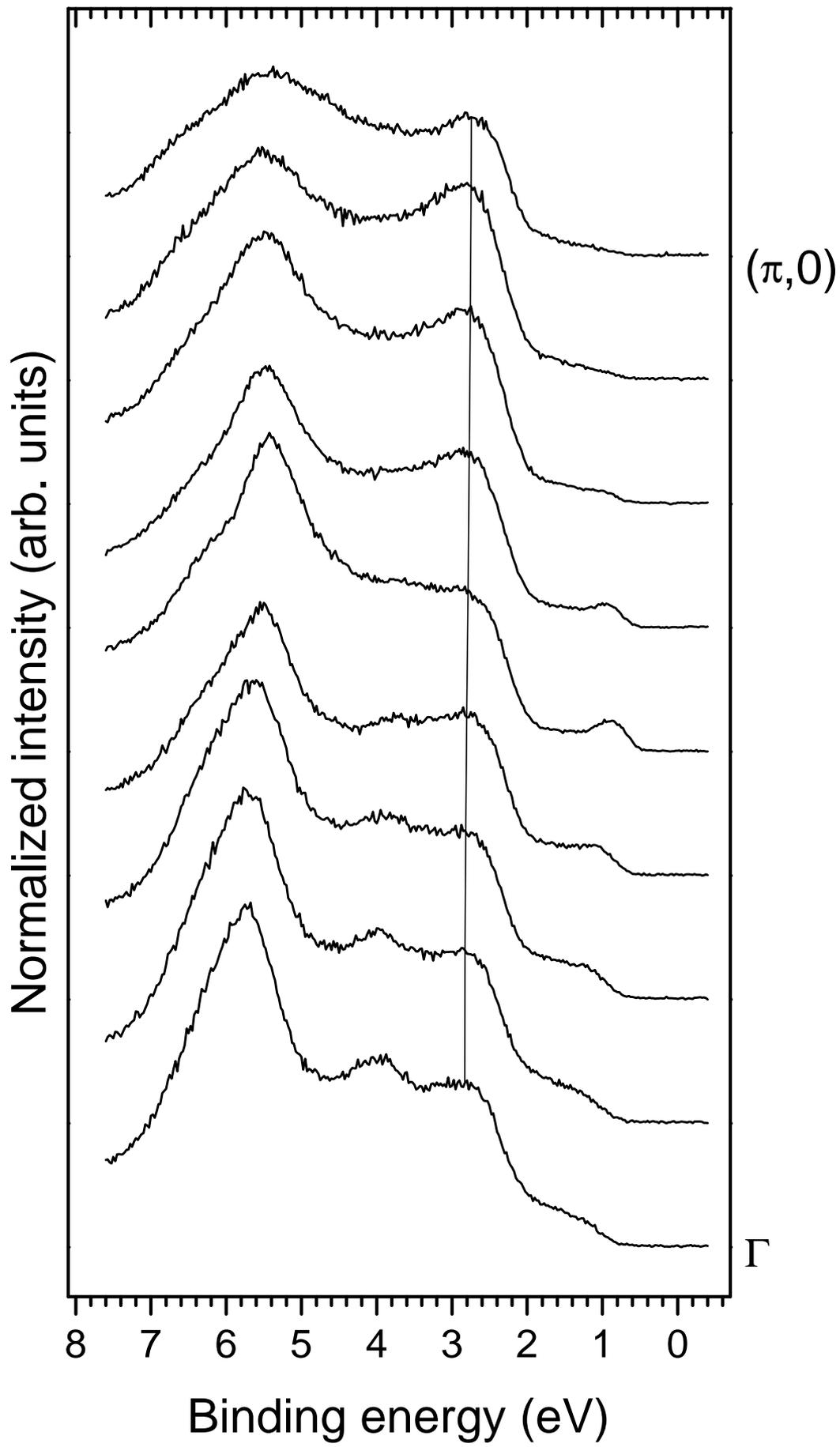
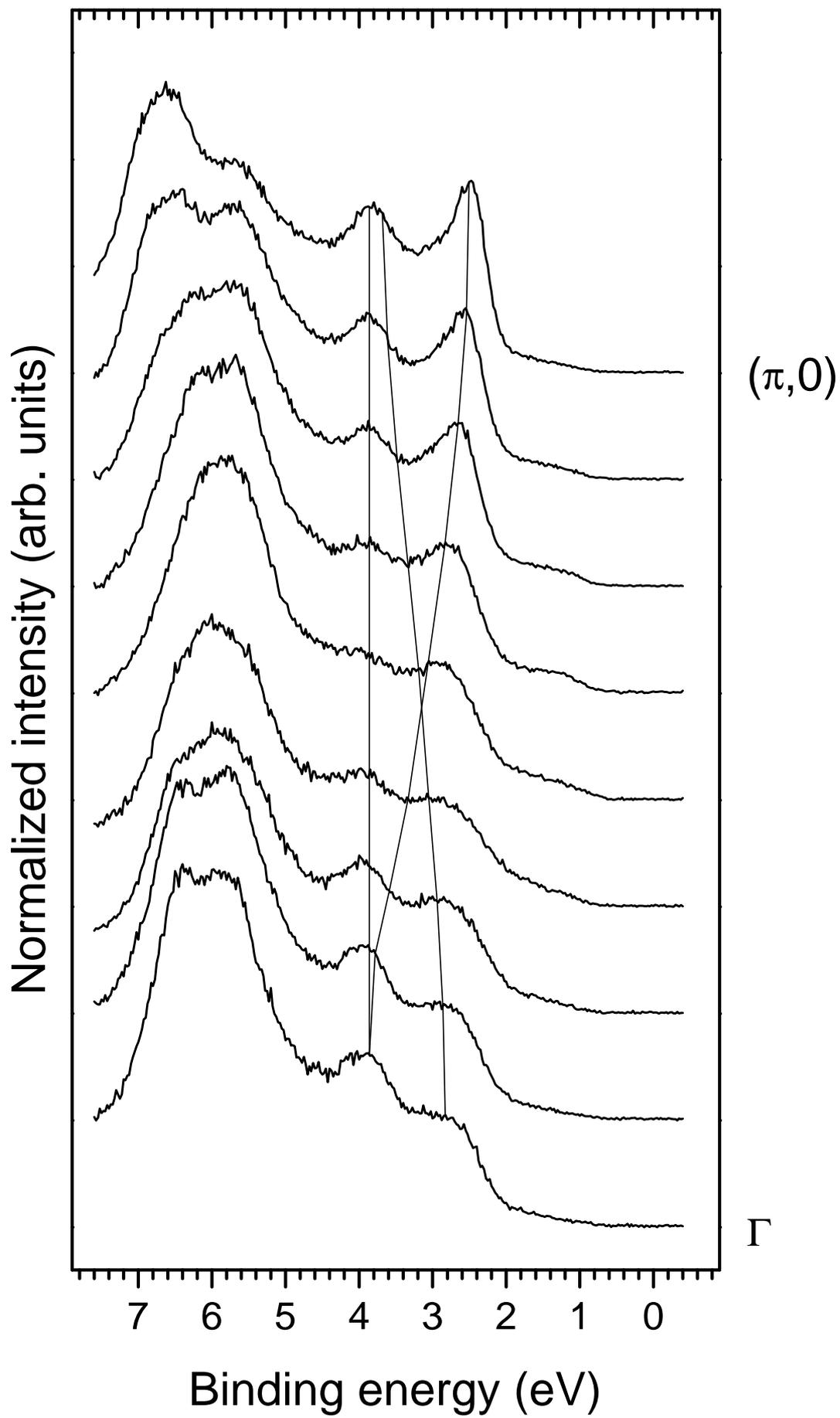
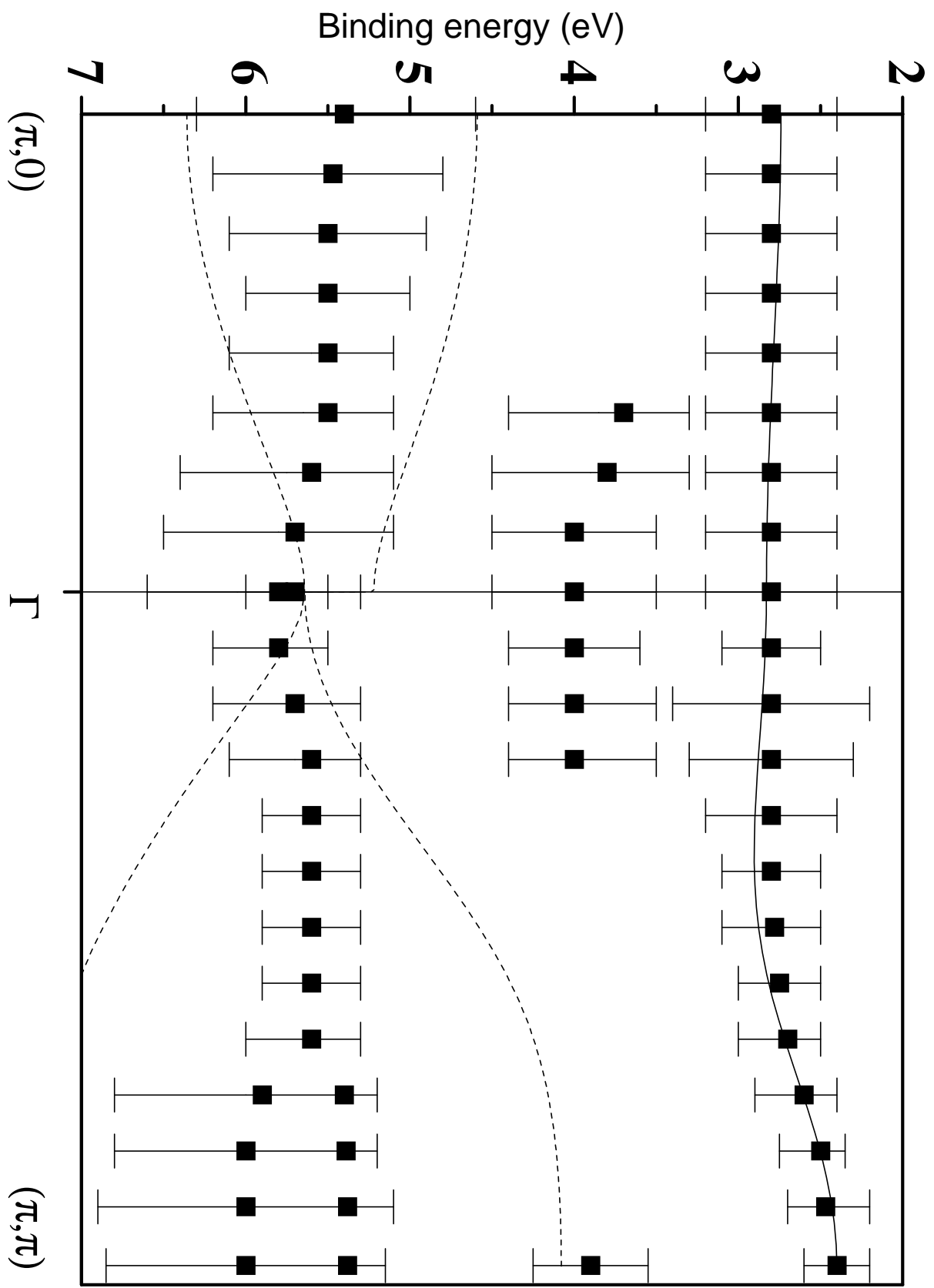
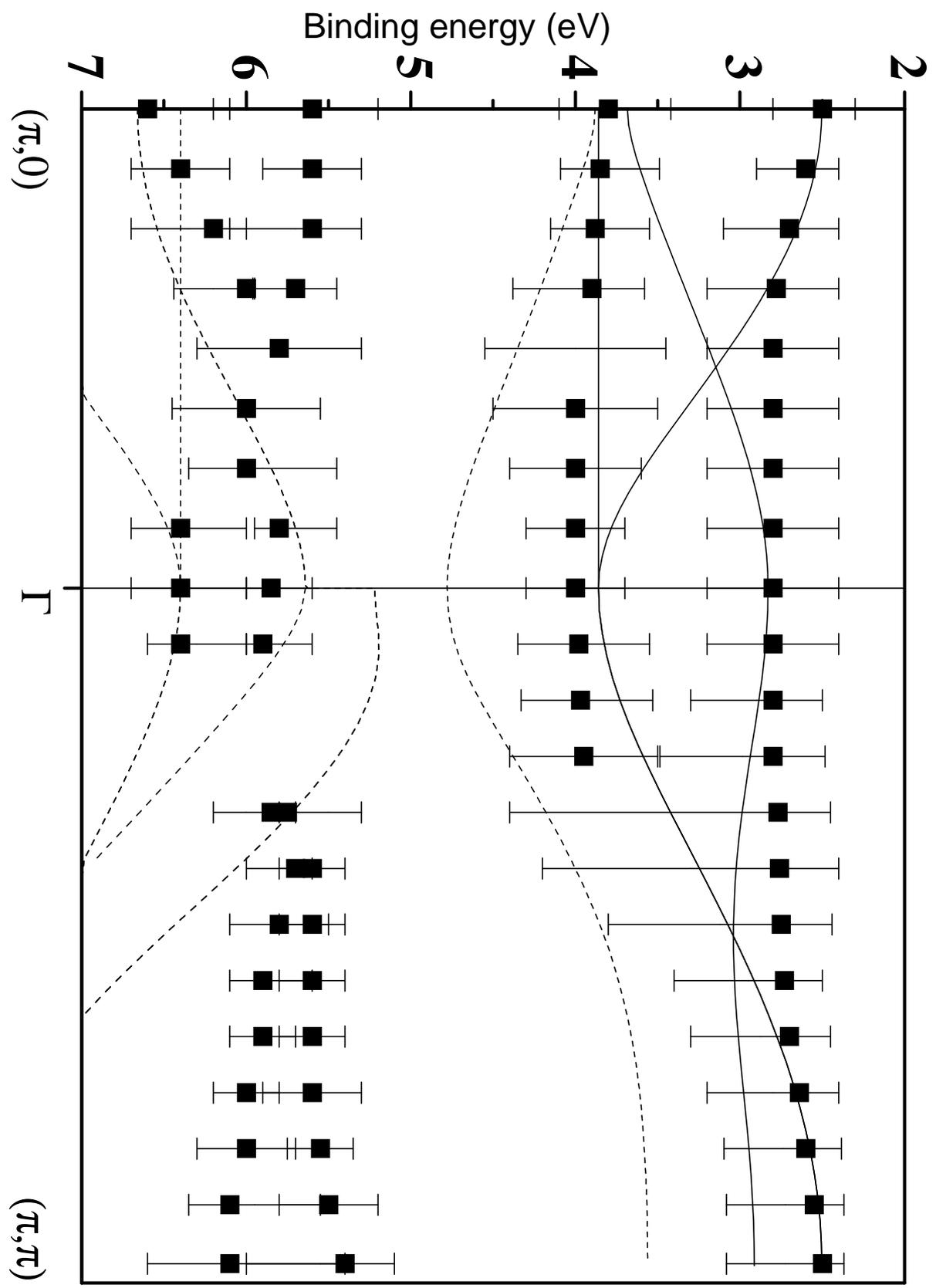
\end{document}